\documentclass[review]{elsarticle} 

\usepackage{lineno}
\modulolinenumbers[5]

\usepackage{color}

\usepackage{mathtools}
\usepackage{bm}
\usepackage{amssymb}
\usepackage{subcaption}

\journal{Control Engineering Practice}

\begin{document}

\begin{frontmatter}

\title{Robust converter-fed motor control based on active rejection of multiple disturbances\tnoteref{mytitlenote}}
\tnotetext[mytitlenote]{\textcolor{black}{The work has been partially supported by “the Fundamental Research Funds for the Central Universities” project no.~21620335.}}


\author[JNU]{Rafal~Madonski\corref{mycorrespondingauthor}}
\ead{rafal.madonski@jnu.edu.cn}
\cortext[mycorrespondingauthor]{Corresponding author}
\author[PUT]{Krzysztof~{\L}akomy}
\author[SER]{Momir~Stankovic}
\author[CSU]{Sally~Shao}
\author[SEU]{Jun~Yang}
\author[SEU]{Shihua~Li}

\address[JNU]{Energy Electricity Research Center, International Energy College, Jinan University, 206 Qianshan Road, Zhuhai, Guangdong, 519070 P.~R.~China}
\address[PUT]{Institute of Automation and Robotics, Poznan University of Technology, Poznan, Poland}
\address[SER]{Military Academy, University of Defense, Belgrade, Serbia}
\address[CSU]{Department of Mathematics, Cleveland State University, Cleveland, OH, USA}
\address[SEU]{School of Automation, Southeast University, Key Laboratory of Measurement and Control of CSE, Ministry of Education, Nanjing, P.~R.~China}

\begin{abstract}
In this work, an advanced motion controller is proposed for buck converter-fed DC motor systems. The design is based on an idea of active disturbance rejection control (ADRC) with its key component being a custom observer capable of reconstructing various types of disturbances (including complex, harmonic signals). A special formulation of the proposed design allows the control action to be expressed in a concise and practically appealing form reducing its implementation requirements. The obtained experimental results show increased performance of the introduced approach over conventionally used methods in tracking precision and disturbance rejection, while keeping similar level of energy consumption. A stability analysis using theory of singular perturbation further supports the validity of proposed control approach.
\end{abstract}


\end{frontmatter}

\section{Introduction}

With the rapid development of power electronics devices, the use of DC-DC buck converters has become an interesting alternative to linear regulators for DC motor control~\cite{SiraBookDCDC}. The DC-DC buck converter-DC motor combination offers a ''smooth start'' of the drive, which mitigates the unwanted effects in the armature circuit. In order to achieve high performance of motion control, this start has to be realized in engineering practice in the inevitable presence of various sources of uncertainties \cite{buckconverters-CEP,SiraDCDC}, including unknown load torques; ii) various parametric uncertainties including those coming from external voltage source, winding resistance, and load resistance; as well as iii) unmodeled dynamics of the sensing device, equivalence series resistance, and direct current resistance in the converter system. This amount of uncertainty in conventional converter-fed motor systems often goes beyond the capability of a standard industrial off-the-shelf PID controller~\cite{Latin}. As a result, numerous advanced control techniques have been proposed to date, including differential flatness-based controller~\cite{Flores2004a}, neuroadaptive backstepping method~\cite{Nizami}, generalized proportional-integral control~\cite{Flores2004c}, hierarchical cascade-like scheme~\cite{Ortigoza3}, nonlinear adaptive controller~\cite{adapCLF}, and most recently sliding mode control~\cite{ArshadTIMC}.

The high performance control of converter-fed motors has been recently investigated from the perspective of active disturbance rejection control (ADRC). This general control concept has been introduced in~\cite{Han-fromPID} and its latest developments has been recently summarized in surveys~\cite{TIE2019p2,obs-survey2,RMsurvey}. The interest in this particular class of techniques came from their recent applications to various motion~\cite{eleCylinder,CEPadaptRESO,KLspatial}, power~\cite{WeiMagLev,fuelcellpowerplant-CEP,powerAplic-CEP}, and process~\cite{Sunli02CEP,SimoneS,disturbancedecoupling-CEP} control problems as well as successful transition of ADRC from academia to industry through its incorporation in embedded motion control chips from Texas Instruments (called InstaSPIN\footnote{http://www.ti.com/microcontrollers/c2000-real-time-control-mcus/applications/instaspin.html (last visit: May 2020)}). Through the utilization of profound implications of the integrator chain form \cite{TDconcept,Gao-centrality} and the concept of real-time \textit{total disturbance} reconstruction and attenuation, several ADRC schemes have been shown to give promising results in governing converter-driven motor systems, e.g.~\cite{WuHanTIE,DCDCDCadrc,SiraDCDCgpi}.

The problem investigated in this work concerns two major limitations of the currently available ADRC methods for converter-fed DC motors. The first one comes from their expression in 2DOF output-based form (in which one degree-of-freedom deals with real-time disturbance reconstruction and rejection and the other for governing the resultant simplified plant model). Although such topology offers important robust and adaptive features, its hardware implementation may be problematic in the cases of motor angular velocity tracking, which nominally use high-order plant models for controller design. In scenarios where a reduced-order observer cannot be used (due to limited plant modeling/sensing capabilities), the conventional ADRC structure requires the availability of multiple consecutive time-derivatives of the output signal and the reference signal in order to synthesize the controller~\cite{MMsygnref,MadonskiCEPerror}. It is rarely the case in industrial applications of converter-fed drives that analytical forms of all these signals are available or that these signals are directly measurable.

The second limitation investigated in this work is the often use of disturbance observers in ADRC that are based solely on a polynomial representation of the total disturbance. This makes conventional ADRC-based approaches only practically capable of handling slowly varying disturbances~\cite{TIE2019p1}. Consequently, the conventional polynomial model limits the abilities to capture the fast varying harmonic disturbances~\cite{TIE2019p2}, which are common in converter-driven DC motor systems. For example,  harmonic currents can cause adverse effects in power systems such as overheating, interferences in sensitive communication equipment, capacitor blowing, motor mechanical vibration, excessive neutral currents, or resonances with the power grid. The harmonic disturbances thus negatively influence the tracking performance, justifying the need for their mitigation through the governing scheme.

Motivated by the above limitations, a new ADRC design is proposed in this work. Its goal is to retain the desired capabilities of standard 2DOF output-based ADRC, while minimizing its disadvantages related to impractical assumptions about signal availability and limited capabilities for attenuating complex multifarious disturbances. The proposed design utilizes a special state transformation and a dedicated observer. The transformation allows the control action to be expressed in a concise practically appealing form reducing its implementation time and requirements, thus making it more easily deployable in various industrial control platforms. The observer is used to virtually decompose the acting lumped disturbance into polynomial-like and sinusoidal-like signals. These two acting disturbance models allow to represent a majority of multifarious disturbances with a satisfactory level of approximation, then simultaneously reconstruct them with a single observer, and finally compensate their effect on the governed output signal in real-time.



In order to verify the efficacy of the proposed design, several experiments are conducted on a laboratory platform to evaluate the performance of the introduced approach. The proposed control technique is compared with some conventional solutions using several criteria like tracking precision, disturbance rejection, and energy consumption. Furthermore, the stability of the proposed control system is proved using singular perturbation theory.


\section{Preliminaries}

\subsection{Simplified plant model}

Following~\cite{SiraBookDCDC}, dynamic model of a buck converter can be expressed as:
	\begin{equation}
	    \begin{cases}
		    L\frac{di}{dt} = -v+Eu,\\
		    C\frac{dv}{dt} = i-\frac{v}{R},
	    \end{cases}
		\label{eq:convermode}
	\end{equation}
in which $u \in [0,1]$ denotes duty-cycle, $i$[A] represents current across the inductor, $R$[$\Omega$] is the load resistance, $C$[F] is the output filter capacitance, $E$[V] denotes the external voltage, $L$[H] represents the input circuit inductance, and $v$[V] is the output voltage.

On the other hand, a permanent magnet DC motor (assuming nonzero armature inductance) can be modeled as:
	\begin{equation}
    	\begin{cases}
    		L_a\frac{di_a}{dt} = v_a-R_a i_a-k_e \omega,\\
    		J\frac{d\omega}{dt} = k_mi_a-b\omega-\tau,
    		\label{eq:motormode}
    	\end{cases}
	\end{equation}
where $\omega=\frac{d\theta}{dt}$[rad/s] represents the motor shaft angular velocity subjected to load torque $\tau$[Nm], $\theta$[rad] denotes the motor shaft angular position, $J$[kg$\cdot$m$^2$] is the inertia of rotor and load, $b$[Nm-s/rad] represents the friction coefficient, $v_a$[V] is the motor armature voltage, $i_a$[A] is the armature current, $k_e$[Vs/rad] is the counter-electromotive force constant, $k_m$[N-m/A] is the motor torque constant, $L_a$[H] is the armature inductance, and $R_a$[$\Omega$] is the armature resistance.

\begin{figure}[t]
	\centering
		\includegraphics[width=\textwidth]{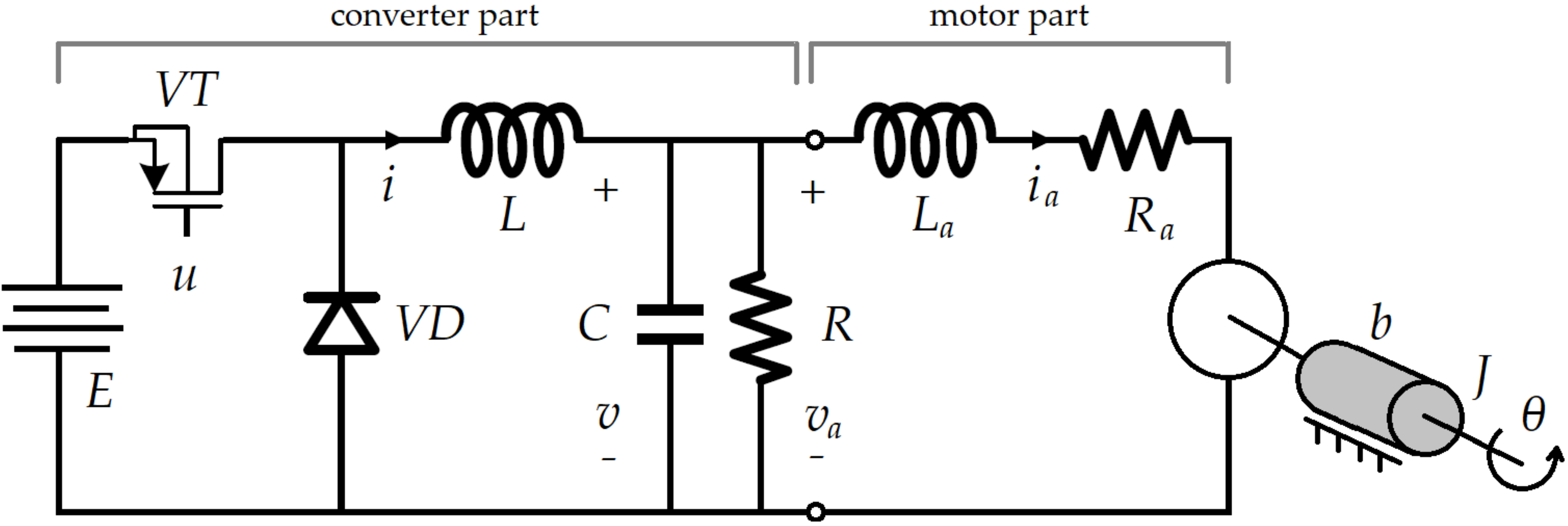}
	\caption{The circuit diagram of the considered converter-fed motor system.}
	\label{fig:DCDC_DC}
\end{figure}

A combination of the converter and motor parts forms the considered converter-fed motor system, seen Fig.~\ref{fig:DCDC_DC}, that can be modeled as:
	\begin{equation}
			\begin{cases}
			C\frac{dv}{dt} = i-\frac{v}{R}-i_a,\\
			L\frac{di}{dt} = -v+Eu,\\
			L_a\frac{di_a}{dt} = v-R_a i_a-k_e \omega,\\
			J\frac{d\omega}{dt} = k_m i_a-b_m\omega-\tau.
			\label{eq:DCDCDC}
			\end{cases}
	\end{equation}
The definition of a state vector  $\bm{x}=[x_1~x_2~x_3~x_4]^\top\triangleq[i~v~i_a~\omega]^\top\in\mathcal{X}\subset\mathbb{R}^4$, allows to express the above system in state-space form as:
	\begin{equation}
	    \begin{cases}
		    \bm{\dot{x}} = \bm{Ax} + \bm{B}_u u + \bm{B}_d \tau, \\
		    \omega = \bm{C}\bm{x},
		\end{cases}
		\label{eq:modelmodel}
	\end{equation}
	where $u$ is the system input, $\omega$ is the system output, while matrices $\bm{C}=[0~0~0~1]\in\mathbb{R}^{1\times4}$,
	$\bm{B}_d=[0~0~0~-1/J]^\top\in\mathbb{R}^{4\times1}$, $\bm{B}_u=[\frac{E}{L}~0~0~0]^\top\in\mathbb{R}^{4\times1}$ and
	\[ \bm{A}=
\begin{bmatrix}
    0 & -\frac{1}{L} & 0 & 0\\
    \frac{1}{C} & -\frac{1}{CR} & -\frac{1}{C} & 0\\
    0 & \frac{1}{L_a} & -\frac{R_a}{L_a} & -\frac{k_e}{L_a}\\
		0 & 0 & \frac{k_m}{J} & -\frac{b_m}{J}\\
\end{bmatrix}\in\mathbb{R}^{4\times4}.
\]

\begin{itemize}
    \item[A1.] Load torque disturbance $\tau$, and its consecutive time-derivatives are bounded, in the sense that there exists a constant $r_\tau\in\mathbb{R}_+$, such that:
    \begin{equation}
        \forall_{t\geq0}\max\left\{\tau(t),\dot{\tau}(t),...,\tau^{(3)}(t)\right\}<r_\tau.\nonumber
    \end{equation}
    \item[A2.] System output signal $\omega$[rad/s] is the only available quantity in the considered control system.
\end{itemize}

\textit{Remark 1.} Several approximations have been made in~\eqref{eq:modelmodel}. Such system model will be used in further considerations regardless due its appealing practical simplicity and the fact that a robust controller will be designed.

\subsection{Control objective}
\label{sect:OBJ}

In this article, the focus is on trajectory tracking motion task of a motor shaft angular velocity $\omega(t)$. A smooth reference motor shaft velocity is defined as $\omega_d(t)$[rad/s]$\in\mathbb{R}^{+}$ and satisfies following practical assumptions:
\begin{itemize}
    \item[A3.] The reference velocity itself, together with its consecutive time-derivatives $\left(\dot{\omega}_d,\ldots,\omega^{(4)}_{d}\right)$, is bounded at any time instant $t$, i.e., there exists a positive constant $r_{\omega_d}\in\mathbb{R}_+$ such that $\forall_{t\geq0}\max\{\omega_d(t),\dot{\omega}_d(t),\ldots,\omega^{(4)}_{d}(t)\}<r_{\omega_d}$.
	\item[A4.] Reference $\omega_d$ is not known in advance, meaning that neither its analytical form, its future values, nor its consecutive reference time-derivatives 
	are available, thus they cannot be utilized in controller/observer synthesis.
\end{itemize}


The control objective is to design a control signal $u(t)$ that will satisfy a precise following of the desired trajectory $\omega_d(t)$ by the motor shaft angular velocity $\omega(t)$, despite the presence of various uncertainties corresponding to system parameters variations, external harmonic interferences, unknown load torques, and unavailability of measuring time-derivatives of $\omega$. Assuming practicality of the desired result, the goal is to make the tracking error $|e(t)|\triangleq |\omega_d(t)-\omega(t)|$ ultimately bounded by some small (but practically acceptable) number $r_e\in\mathbb{R}_+$, i.e. $\limsup_{t\geq0}|e(t)|<r_e$.



\subsection{Flatness-based model representation}


In order to express the plant mathematical model in a causal input-output integral chain perturbed with a matched, input-additive, lumped disturbance, for which most ADRC-based control schemes are most conveniently derived for, the flatness property of the considered system is utilized. Applying the notation from~\cite{flatSira}, the SISO system~\eqref{eq:DCDCDC}, expressed generally as $\dot{\bm{x}}=g(\bm{x},u,\tau)$, is differentially flat if there exists a scalar output $\zeta = \Phi(\bm{x})\in\mathbb{R}$, called \textit{flat} output, which allows a differential parametrization of state ($\bm{x}$) and input ($u$) in form of:
	\begin{equation}
		\bm{x} = \Psi_x\left(\zeta,\dot{\zeta},\ddot{\zeta},\zeta^{(3)}\right),~u = \Psi_u\left(\zeta,\dot{\zeta},\ldots,\zeta^{(4)}\right).\label{eq:flat2}
	\end{equation}

Recalling that for linear systems the concept of flatness is equivalent to a controllability property, one can easily verify the flatness of~\eqref{eq:DCDCDC}. We thus compute its controllability matrix as:
\color{black}
\begin{equation}
\bm{Q}_C=
\begin{bmatrix}
    \bm{B}_u\\
    \bm{AB}_u\\
		\bm{A^2B}_u\\
    \bm{A^3B}_u\\
\end{bmatrix}^\top=
\begin{bmatrix}
    \frac{E}{L} & 0 & -\frac{E}{CL^2} & -\frac{E}{C^2L^2R}\\
    0 & \frac{E}{CL} & -\frac{E}{C^2LR} & -\frac{E}{C^2L^2R^2L_a}a_{24}\\
    0 & 0 & \frac{E}{CLL_a} & -\frac{E}{CLL_a}a_{34}\\
		0 & 0 & 0 & \frac{E}{CJLL_a}k_m
\end{bmatrix},
\end{equation}
\color{black}
with $a_{24}=CLR^2-LL_a+CR^2L_a$, and $a_{34}=EL_a+ECRR_a$.

The flat output is given by the inner product of the last row of the inverse of the Kalman controllability matrix $\bm{Q}_C$, which yields:
\begin{equation}
	\zeta = \overbrace{[0~0~0~1]}^{\bm{C}}\bm{Q}_C^{-1}\bm{x} = \left(\frac{Ek_m}{CJLL_a}\right)^{-1}x_4=\frac{1}{\bm{CA}^3\bm{B}_u}x_4 = b_0^{-1}x_4,
\end{equation}
where $b_0=\frac{Ek_m}{CJLL_a}$. Using flat output $\zeta$, a transformation of~\eqref{eq:DCDCDC} to the \textit{control normal form} gives:
	\begin{align}
		\bm{\nu} &= [\nu_1~\nu_2~\nu_3~\nu_4]^\top\nonumber\\
		&=\left[\zeta~\dot{\zeta}~\ddot{\zeta}~\zeta^{(3)}\right]^\top\nonumber\\
		&=
    \left[\bm{C}^\top,~\bm{A}^\top\bm{C}^\top,~\left(\bm{A}^2\right)^\top\bm{C}^\top,~\left(\bm{A}^3\right)^\top\bm{C}^\top\right]^\top
		\bm{x}\nonumber\\
		&=\bm{T}\bm{x}\in\mathcal{N}\subset\mathbb{R}^4,
	\end{align}
which results in a transformed system: $\dot{\bm{\nu}} = \bm{TAT}^{-1}\bm{\nu} + \bm{TB}_u u + \bm{TB}_d \tau$. The differential parametrization of the input is given by solving the last row of transformed system $\dot{\bm{z}}$ for input $u$, that yields $u=\Psi_u(\bm{\nu},\dot{\nu}_4)$. Similarly, with the inverse $\bm{x}=\bm{T}^{-1}\bm{z}:\mathcal{N}\rightarrow\mathcal{X}\subset\mathbb{R}^4$, one can establish the relations in~\eqref{eq:flat2}, which can be straightforwardly utilized in the ADRC design. Utilization of the flatness property allows to rewrite system~\eqref{eq:modelmodel} as:
\begin{align}
	\omega^{(4)} &= \bm{Cx}^{(4)}\nonumber\\
	&= \bm{CA}^3\left(\bm{Ax} + \bm{B_u}u + \bm{B_d}\tau\right) + \bm{CA}^2\bm{B_d} \dot{\tau}\nonumber\\
	&+ \bm{CAB_d} \tau^{(2)} + \bm{CB_d} \tau^{(3)},
    \label{eq:4thordsyst}
\end{align}
which has explicit form of coordinates transformation:
	\begin{equation}
\bm{x} = \begin{bmatrix}
    \bm{C} \\
    \bm{CA} \\
		\bm{CA^2} \\
		\bm{CA^3}
\end{bmatrix}^{-1} \left(\bm{z}-\begin{bmatrix}
    0 \\
    \bm{CB}_d \\
		\bm{CAB}_d \\
		\bm{CA^2}\bm{B}_d
\end{bmatrix}\tau-\begin{bmatrix}
    0 \\
    0 \\
		\bm{CB}_d \\
		\bm{CA}\bm{B}_d
\end{bmatrix}\dot{\tau}-\begin{bmatrix}
    0 \\
    0 \\
		0 \\
		\bm{CB}_d
\end{bmatrix}\ddot{\tau}\right).\nonumber
	\end{equation}

Further rearrangement of~\eqref{eq:4thordsyst} results in a simplified model:
\begin{align}
	\omega^{(4)} = \tilde{F}\left(t,\bm{x},\tau,\dot{\tau},\ddot{\tau},\tau^{(3)},u\right)  + \hat{b}_0u,~~\hat{b}_0\neq 0,
	\label{eq:standmodeladrc}
\end{align}
where $\hat{b}_0\approx b_0$ is a rough approximation of the system input gain (inevitably subject to parametric mismatch) and term:
\begin{align}
    \tilde{F}(t,\cdot)&=\bm{CA}^4\bm{x}+\bm{CA}^3\bm{B}_d\tau+\bm{CA}^2\bm{B}_d\dot{\tau}+\bm{CAB}_d\ddot{\tau} \nonumber \\
    &+\bm{CB}_d\tau^{(3)}+\bm{CA}^3\bm{B}_uu-\hat{b}_0u,
\end{align}
denotes the plant aggregated ''total disturbance''. In general, it comprises of endogenous uncertainties (representing parametric uncertainties and either unknown or purposefully neglected state-depended terms, whether of linear or nonlinear nature) and exogenous disturbances (represented by unknown, unstructured, disturbances signals - possibly of harmonic nature, as well as noises, etc.).

\textit{Remark 2.} The justification of such compact and constantly updated plant description~\eqref{eq:standmodeladrc}  has been thoroughly validated in the area of active disturbance rejection~\cite{Wenchao-NTV,TDconcept2,TDconcept} from both theoretical and practical points of view~\cite{TIE2019p2,obs-survey2,RMsurvey}. In particular, the treatment of state-dependent and control signal-dependent terms as part of the total disturbance has been already addressed in the literature (see Sect.~III.C in~\cite{GESOmismatchh}).

\subsection{Conventional ADRC}
\label{sec:standaaa}

The conventional active disturbance rejection control solution, as seen in~\cite{Gao-pole-placement} for~\eqref{eq:standmodeladrc} would be a two-stage governing action in form of:
\begin{equation}
	u \triangleq
	\frac{1}{\hat{b}_0}\Bigg(\underbrace{\omega_d^{(4)}+\sum_{i=0}^{3}k_ie^{(i)}}_{\tilde{u}_0}-\hat{\tilde{F}}(t,\cdot)\Bigg),
	\label{eq:standardADRCu}
\end{equation}
where the first stage is on-line disturbance estimation and rejection $(\hat{\tilde{F}}(t,\cdot))$ and the second stage is control objective realization $(\tilde{u}_0)$ with an output feedback plus feed-forward controller, designed for an idealized disturbance-free system $\omega^{(4)}=u$ (cf.\eqref{eq:standmodeladrc}), which behavior is shaped by proper selection of controller coefficients $k_i>0$ for $i\in\{0,...,3\}$.

An implementation of the conventional ADRC governing action~\eqref{eq:standardADRCu} into system~\eqref{eq:standmodeladrc} reveals the limitations with its practical application as several higher-order terms are required to be available,
which in the case of the considered converter-fed system violates assumptions A2 and A4. The other limitation is the often used polynomial model for reconstructing the total disturbance term, whereas $\tilde{F}$ in the considered system consists of multiple-type disturbances, including complex harmonic ones.

In order to address the above limitations, while not loosing the robustness of tracking performance offered by~\eqref{eq:standardADRCu}, a new ADRC design for converter-driven DC motors is introduced next.


\section{Proposed ADRC structure}




\subsection{Control task reformulation}

Recalling the definition of $e(t)$, one can rewrite~\eqref{eq:standmodeladrc} in  error-domain as:
\begin{equation}
	e^{(4)} \triangleq \omega_d^{(4)} - \omega^{(4)} \stackrel{\mathclap{\mbox{\scriptsize{\eqref{eq:standmodeladrc}}}}}{=} \omega_d^{(4)} - \tilde{F}\left(t,\cdot\right)  - \hat{b}_0u.
	\label{eq:eftylda}
\end{equation}
Furthermore, term $\sum_{i=1}^{3}k_ie^{(i)}$ can be added to both sides of~\eqref{eq:eftylda} giving:
	\begin{equation}
		e^{(4)}+\sum_{i=1}^{3}k_ie^{(i)}=\underbrace{\sum_{i=1}^{3}k_ie^{(i)}+\omega_d^{(4)} -\tilde{F}\left(t,\cdot\right)}_{F(t,\cdot)}-\hat{b}_0u,
		\label{eq:MMerr2}
	\end{equation}
where $F$ is now the total disturbance for the modified system model in error-based form with partially incorporated closed-loop dynamics (cf.\eqref{eq:standmodeladrc}).


With the introduced alternative system description in~\eqref{eq:MMerr2}, a control signal can be designed as (cf.\eqref{eq:standardADRCu}):
\begin{equation}
	u \triangleq \frac{1}{\hat{b}_0}\left(u_0+\hat{F}\right) = \frac{1}{\hat{b}_0}\left(k_0 e + \hat{F}\right),~u_0 = k_0 e,
	\label{eq:MMu}
\end{equation}
which when applied to~\eqref{eq:MMerr2} gives the following dynamics:
\begin{equation}
	e^{(4)}+k_3e^{(3)}+k_2\ddot{e}+k_1\dot{e}+k_0e = F-\hat{F},
	\label{eq:errordynaam}
\end{equation}
with its behavior in accordance to user-defined gains $k_i>0$ and $i\in\{0,..,3\}$.

\textit{Remark 3.} Note that $u_0$ is chosen in the proposed control action~\eqref{eq:MMu} as a proportional controller, which trivializes the entire control synthesis in comparison to conventional solutions~\eqref{eq:standardADRCu}. The otherwise unavailable target time-derivatives, needed for fourth-order integral chain stabilization, are now incorporated as a part of $F$, to estimated by an observer (designed next).

\subsection{Complex disturbance reconstruction}

It becomes clear from~\eqref{eq:MMu} that the quality of control task realization in practice will be a function of total disturbance estimation quality. Additionally, to address the considered limitation of conventional ADRC design with effective harmonic disturbance estimation, we first adopt our general methodology~\cite{RESOs,MSSPmad,MomirIJC} and express $F$ from~\eqref{eq:MMerr2} as a sum:
\begin{equation}
F(t)=F_{p}(t)+F_{w}(t,\omega_r)+F_r(t),
\label{eq:sumasinnspoly}
\end{equation}
where $F_{p}(t)=c_0+c_1t+c_2t^2+\dots+c_{m-1}t^{(m-1)}$ is a polynomial disturbance component, $F_{w}(t,\omega_r)=a_1\sin(\omega_r t)+a_2\cos(\omega_r t)$ is a sinusoidal disturbance, while $F_r(t)$ includes disturbances that do not match models $F_p(t)$ or $F_w(t,\omega_r)$.
A following model of an unforced oscillator can be used to represent the behavior of $F_w(t,\omega_r)$:
\begin{equation}
\ddot{F}_w(t,\omega_r)+\omega_r^2F_w(t,\omega_r)=0,
\label{equ20}
\end{equation}
where $\omega_r$ is its resonant frequency.

\textit{Remark 4.}
According to~\cite{RESOs,MomirIJC}, we can design a disturbance observer that allows to fully compensate $F_p(t)$ and $F_w(t,\omega_r)$ (assuming perfect knowledge of $\omega_r$), and estimate $F_r(t)$ with an arbitrary high precision depending on the observer specific structure and design parameters.


In this particular work, we would like to be able to completely reject a constant disturbance $F_{p}(t)=c_0$ (i.e. $m=1$) and one harmonic component $F_w(t,\hat{\omega}_r)$, where $\hat{\omega}_r$ is the estimated pulsation of the real $\omega_r$. To achieve that, we need to define the extended state for system~\eqref{eq:MMerr2} in a form $\bm{z} = [z_1 \ ... \ z_7]^\top\triangleq[e \ \dot{e} \ \ddot{e} \ e^{(3)} \ F \ F_w \ \dot{F}_w]^\top\in\mathbb{R}^7$. Such extended state representation leads to a controllable and observable system expressed in state-space as:
\begin{equation}
\begin{cases}
		\dot{\bm{z}} = \bm{\bar{A}z} - \hat{\bm{b}}_0 (u-u_0) + \bm{h}\left(\dot{F}_r+\dot{F}_w(\cdot,\omega_r)-\dot{F}_w(\cdot,\hat{\omega}_r)\right),\\
		e = \bm{c}\bm{z},
		\label{eq:nowyyysystSS}
\end{cases}
\end{equation}
where
\[ \bm{\bar{A}}=
\begin{bmatrix}
    0 & 1 & 0 & 0 & 0 & 0 & 0\\
		0 & 0 & 1 & 0 & 0 & 0 & 0\\
		0 & 0 & 0 & 1 & 0 & 0 & 0\\
    -k_0 & -k_1 & -k_2 & -k_3 & 1 & 0 & 0\\
		0 & 0 & 0 & 0 & 0 & 1 & 0\\
		0 & 0 & 0 & 0 & 0 & 0 & 1\\
    0 & 0 & 0 & 0 & 0 & -\hat{\omega}_r^2 & 0\\
\end{bmatrix}\in\mathbb{R}^{7\times7},
\hat{\bm{b}}_0=
\begin{bmatrix}
    0\\
    0\\
		0\\
		\hat{b}_0\\
		0\\
		0\\
		0
\end{bmatrix}\in\mathbb{R}^{7\times1}
\]
$\bm{h}=\left[0~0~0~0~1~0~0\right]^\top\in\mathbb{R}^{7\times1}$ and $\bm{c}=[1~0~0~0~0~0~0]\in\mathbb{R}^{1\times7}$.

Now, a resonant extended state observer (RESO) for system~\eqref{eq:nowyyysystSS} can be derived in error-based form to reconstruct the state vector (including the unknown total disturbance $F$) based solely on the already available input ($u$) and output ($e$) signals:
\begin{equation}
\begin{cases}
		\dot{\hat{\bm{z}}} = \bm{\bar{A}}\hat{\bm{z}} - \hat{\bm{b}}_0 (u-u_0) +\bm{l}\left(e-\hat{e}\right),\\
		\hat{e} = \bm{c}\hat{\bm{z}},
\end{cases}
\label{eq:RESOoo}
\end{equation}
%
where $\hat{\bm{z}}$ is the estimate of $\bm{z}$, $\bm{l}=[l_1~...~l_7]^\top\in\mathbb{R}_+^7$ is the vector of observer positive gains, and $\hat{e}$ is the estimate of $e$.

\begin{figure}[t]
	\centering
		\includegraphics[width=\textwidth]{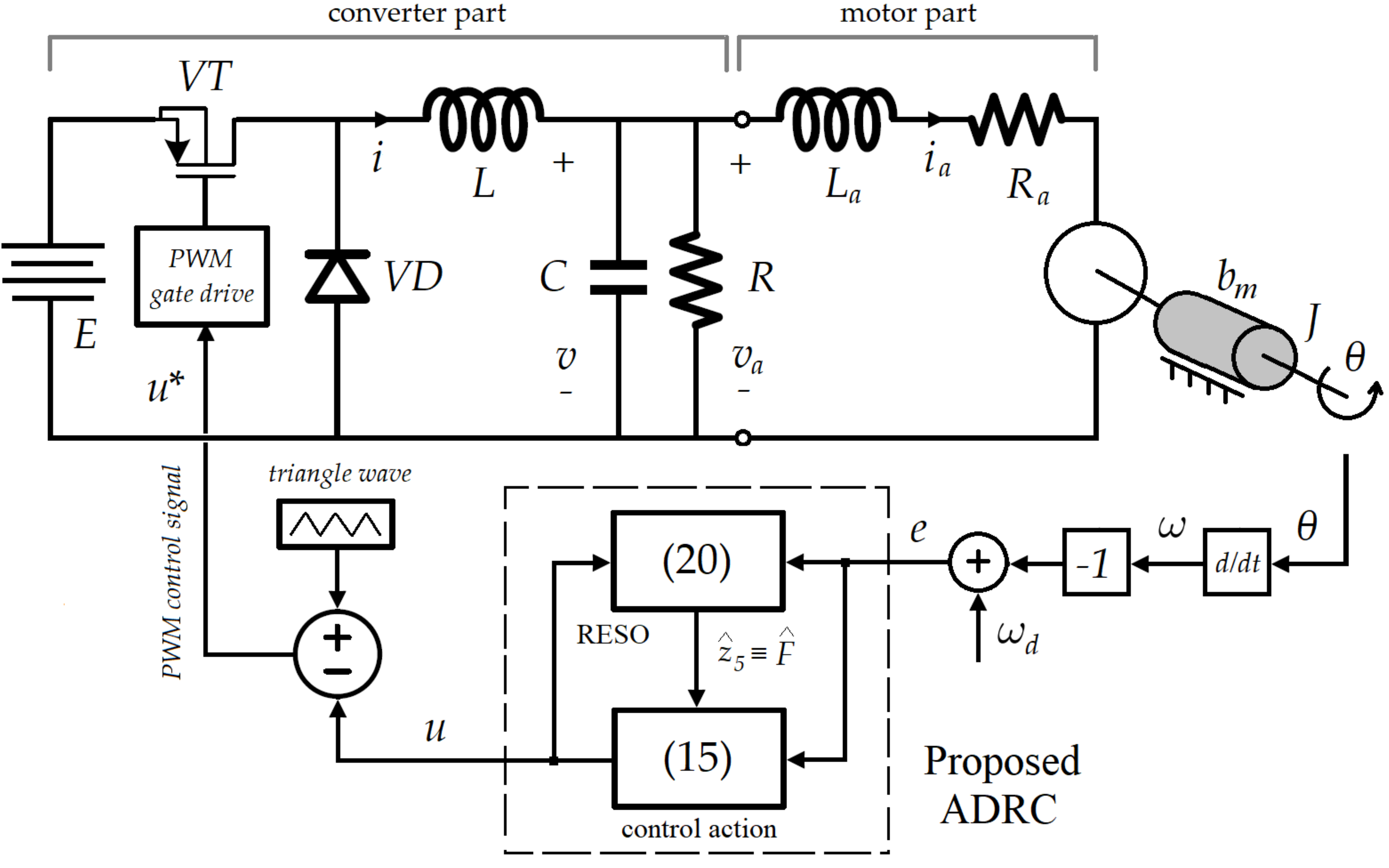}
	\caption{Proposed ADRC with RESO for the converter-fed motor system.}
	\label{fig:DCDCster}
\end{figure}

Thanks to partially incorporating the desired closed-loop dynamics~\eqref{eq:MMerr2}, terms $k_{1}$-$k_{3}$ are now present in $\bm{\bar{A}}$, which means that the control designer can be unburdened from knowing higher-order derivatives of $e$ (see A2 and A4) as they are now estimated by observer~\eqref{eq:RESOoo}. It allows to reduce $u_0$ to just a proportional action~\eqref{eq:MMu}, which significantly reduces the complexity of control synthesis and implementation. This results in several advantages, e.g. fewer sensor needed, reduced implementation time, straightforward interpretation of the control action. A block diagram showing its application to the considered converter-fed motor system is shown in Fig.~\ref{fig:DCDCster}.


\color{black}

\subsection{Stability analysis}
\label{sec:Stabanaalal}

The theoretical investigation concerning proposed ADRC structure with RESO is based on a singular perturbation theory~\cite{Singular}. That is why certain definitions and transformations need to be introduced next that will be later utilized in the stability proof.


First, upon \eqref{eq:nowyyysystSS} and \eqref{eq:RESOoo},  let us consider the dynamics of the observation error $\bm{q}\triangleq\bm{z}-\hat{\bm{z}}$:
\begin{equation}
	\dot{\bm{q}} = \bm{H}\bm{q} + \bm{h}\eta = \left(\bar{\bm{A}}-\bm{lc}\right)\bm{q} + \bm{h}\eta,
	\label{eq:obswerbbb}
\end{equation}
where
\begin{equation}
	\eta=\dot{F_r}+\dot{F}_w(\cdot,\omega_r)-\dot{F}_w(\cdot,\hat{\omega}_r),\nonumber
\end{equation}
and $\bm{H}$ is the estimation error matrix. A practically convenient and widely known pole-placement approach can be applied to parametrize the observer and controller gains selection~\cite{Gao-pole-placement}. We thus propose to design the characteristic equation:
\begin{align}
	\text{det}(\lambda \bm{I}-\bm{H}) &= \lambda^{7}+s_1\lambda^{6}+
	\ldots+s_{6}\lambda+s_{7}, \label{eq:polycoef}
\end{align}
 as:
\begin{align}
\begin{cases}
	s_1 &= \overbrace{k_{3}}^{r_1}+l_1=\beta_1\omega_o\Rightarrow l_1=\beta_1\omega_o-r_1,  \\
	s_2 &= \overbrace{k_{3}l_1+\hat{\omega}_{r}^2+k_{2}}^{r_2}+l_2=\beta_2\omega_o^2\Rightarrow l_2=\beta_2\omega_o^2-r_2,\\
	&~\vdots \\
	 s_{7} &= \overbrace{l_{5}\hat{\omega}_{r}^2}^{r_{7}}+l_{7}=\beta_{7}\omega_o^{7}\Rightarrow l_{7}=\beta_{7}\omega_o^{7}-r_{7}, \end{cases} \label{eq:wspolpoly}
\end{align}
\normalsize
where $r_i$, recalling that $n=4$, are uniquely defined by $k_i=\frac{n!}{i!(n-i)!}\omega_c^{4-i}$ (for $0\leq i \leq n-1$) and $\hat{\omega}_{r}^2$, while parameters $\beta_i = \frac{(n+3)!}{i!(n+3-i)!} $ (for $1\leq i \leq n+3$). Terms $\omega_o, \omega_c>0$ are design parameters that correspond to the observer and controller bandwidths, respectively. It leads to the possibility of decomposition of the estimation error state matrix $\bm{H}$ as:
\begin{align}
\bm{H} &= \left.\begin{bmatrix}
    -\beta_{1}\omega_o & 1 & 0 & 0 & 0 & 0 & 0\\
		-\beta_{2}\omega_o^2 & 0 & 1 & 0 & 0 & 0 & 0\\
    -\beta_{3}\omega_o^3 & 0 & 0 & 0 & 0 & 0 & 0\\
		-\beta_{4}\omega_o^{4} & 0 & 0 & 0 & 1 & 0 & 0\\
		-\beta_{5}\omega_o^{5} & 0 & 0 & 0 & 0 & 1 & 0\\
		-\beta_{6}\omega_o^{6} & 0 & 0 & 0 & 0 & 0 & 1\\
		-\beta_{7}\omega_o^{7} & 0 & 0 & 0 & 0 & 0 & 0
\end{bmatrix}+\right.\nonumber\\
		&+ \begin{bmatrix}
    r_1 & 1 & 0 & 0 & 0 & 0 & 0\\
		r_2 & 0 & 1 & 0 & 0 & 0 & 0\\
    r_3 & 0 & 0 & 0 & 0 & 0 & 0\\
		r_4 & -k_1 & -k_2 & -k_{3} & 1 & 0 & 0\\
		r_{5} & 0 & 0 & 0 & 0 & 1 & 0\\
		r_{6} & 0 & 0 & 0 & 0 & 0 & 1\\
		r_{7} & 0 & 0 & 0 & 0 & -\hat{\omega}_{r}^2 & 0
\end{bmatrix}.\nonumber
\end{align}

Now, by introducing a change of coordinates to~\eqref{eq:obswerbbb} as: $\bm{q}=\bm{\Lambda\xi}$, with $\bm{\Lambda}=\text{diag}\left\{\omega_o^{-7}~\cdots~\omega_o^{-1}~1\right\}\in\mathbb{R}^{7\times7}$, allows to rewrite it as:
\begin{equation}
	\bm{\Lambda}\dot{\bm{\xi}} =
	\bm{H}\bm{\Lambda}\bm{\xi}+\bm{h}\eta,
	\label{eq:SSadj}
\end{equation}
\color{black}where $h=\left[0,0,0,0,0,0,1\right]^\top$. Multiplying both sides of~\eqref{eq:SSadj} by $\Lambda^{-1}$ from the right, one gets:
\begin{equation}
    \bm{\Lambda}\dot{\bm{\xi}}\Lambda^{-1} = \bm{H}\bm{\Lambda}\bm{\xi}\Lambda^{-1}+\bm{h}\eta\Lambda^{-1},
\end{equation}
\color{black}which yields:
\begin{equation}
 \frac{1}{\omega_o}\bm{\dot{\xi}} = \bm{H}_q\bm{\xi}+\frac{1}{\omega_o}\bm{h}\eta~~\Longrightarrow~~\varepsilon\bm{\dot{\xi}} = \bm{H}_q\bm{\xi}+\varepsilon\bm{h}\eta,
\label{eq:omegahh}
\end{equation}
with $\varepsilon=\frac{1}{\omega_o}$, and $\bm{H}_q=\bm{A}_q+\varepsilon \bm{H}_\varepsilon$, where:
\begin{equation}
\bm{A}_q = \begin{bmatrix}
    -\beta_1 & 1 & 0 & 0 & 0 & 0 & 0\\
    -\beta_2 & 0 & 1 & 0 & 0 & 0 & 0\\
		-\beta_3 & 0 & 0 & 1 & 0 & 0 & 0\\
		-\beta_{4} & 0 & 0 & 0 & 1 & 0 & 0\\
		-\beta_{5} & 0 & 0 & 0 & 0 & 1 & 0\\
		-\beta_{6} & 0 & 0 & 0 & 0 & 0 & 1\\
		-\beta_{7} & 0 & 0 & 0 & 0 & 0 & 0
		\end{bmatrix},
\end{equation}

\begin{equation}
\bm{H}_\varepsilon = \begin{bmatrix}
    r_1 & 0 & 0 & 0 & 0 & 0 & 0\\
    \varepsilon r_2 & 0 & 0 & 0 & 0 & 0 & 0\\
		\varepsilon^2 r_3 & 0 & 0 & 0 & 0 & 0 & 0\\
		\varepsilon^{3}r_{4} & -\varepsilon^{2}k_{1} & -\varepsilon k_{2} & -k_{3} & 0 & 0 & 0\\
		\varepsilon^{4} r_{5} & 0 & 0 & 0 & 0 & 0 & 0\\
		\varepsilon^{5} r_{6} & 0 & 0 & 0 & 0 & 0 & 0\\
		\varepsilon^{6} r_{7} & 0 & 0 & 0 & 0 & -\varepsilon\hat{\omega}_{r}^2 & 0
\end{bmatrix}.\nonumber
\end{equation}


Next, the observer error dynamics~\eqref{eq:omegahh} is considered in form of:
\begin{equation}
	\varepsilon\dot{\xi} = \bm{H}_q \xi + \varepsilon h \cdot \eta,
	\label{eq:nowerowneq}
\end{equation}
where $\xi(0) = \xi_0$ represents the initial condition. The above relation represents the so-called initial value problem of the standard singularly perturbed linear system of differential equations. It should be noted~$\bm{H}_q$ and $\varepsilon\bm{H}$ are similar, i.e.,  $\varepsilon\bm{H}=\bm{\Lambda} \bm{H}_q \bm{\Lambda}^{-1}$,
and have same eigenvalues $\lambda_i(\bm{H}_q)=\lambda_i(\bm{H})=-\omega_o$ (see tuning procedure described by \eqref{eq:wspolpoly}), whereas $\lambda_i(\bm{A}_q)=-1$ (for $1\leq i\leq n+3$), which implies $\bm{H}_q$ and $\bm{A}_q$ are Hurwitz. It means that $\hat{\omega}_{r}$ in $\bm{A}$ does not influence the stability of $\bm{H}_q$ in~\eqref{eq:nowerowneq} as long as $\hat{\omega}_{r}$ is finite.

Having introduced the above definitions and transformations, one can compare the proposed ADRC with RESO with the conventional ADRC and its rigorous proof given in~\cite{Gao-stabilitySingPert}. Similar to the conventional ADRC~\cite{Gao-stabilitySingPert}, it can be concluded here for the proposed proposed ADRC with RESO that for the existing derivative of total disturbance $\dot{F}$, if $\left\|\xi_0\right\|=O(\varepsilon)$ i.e. $\xi_0$ lies in its stable initial manifold, the danger of large magnitude transients in~\eqref{eq:nowerowneq} does not exists as $\varepsilon\rightarrow 0$ for $t>0$, hence the system is exponentially stable and uniformly asymptotically stable (for stability within the stable initial manifold for $\xi_0$)\footnote{It should be noted that this is not stability in the Lyapunov sense.}. Hence, by looking at Theorems 3.1 and 4.1 of conventional ADRC stability analysis in~\cite{Gao-stabilitySingPert}, main stability theorem for the proposed ADRC with RESO can be given as follows.


\textbf{Theorem 1}~\textit{If $\left\|\xi_0\right\|=O(\varepsilon)$ and $\eta=\dot{F}_r$ exists, then:}
\begin{itemize}
	\item [(i)] \textit{asymptotic solution of~\eqref{eq:nowerowneq},}
		\begin{equation}
			\varepsilon\dot{\xi} = \bm{H}_q \xi + \varepsilon h \cdot \eta = (\bm{A}_q+\varepsilon \bm{H}_\varepsilon)\xi+\varepsilon h \cdot \eta,~\xi(0)=\xi_0,\nonumber
		\end{equation}
\textit{is uniformly valid for all finite time $L$ with $0 \leq t \leq L < \infty$, and is expressed as:}
	\begin{align}
		\xi(\varepsilon,t) &= \exp\left(\bm{A}_q \frac{t}{\varepsilon}\right)\xi_0+\varepsilon\left\{\bm{A}_q^{-1}h\cdot \eta+\right.\nonumber\\
		&+ \exp\left(\bm{A}_q\frac{t}{\varepsilon}\right) \left(\bm{A}_q^{-1}h\cdot \eta(0)\right)+O\left(\varepsilon^2\right)\\
		&- \left.\bm{H}_0 \int_0^{t} \exp\left[\bm{A}_q \left(\frac{t-s}{\varepsilon}\right)\right]\bm{A}_q^{-1}h\cdot \eta(s)ds\right\};\nonumber
	\end{align}
		\item [(ii)] \textit{there exists $\varepsilon^*>0$ such that for all $\varepsilon\in\left[0,\varepsilon^*\right]$ the leading term of the solution $\xi(\varepsilon,t)$ in the initial layer of system~\eqref{eq:nowerowneq} is exponentially stable;}
		\item [(iii)] \textit{for all $\xi\in \Omega=\left[-\rho_1,\textcolor{black}{\rho_1}\right]^{7}\subset \mathbb{R}^{7}$ with $\textcolor{black}{\rho_1}=\gamma \max\left\{\omega_c,\frac{1}{\omega_o},\left\|\bm{H}_q\right\|\right\}$ for some constant $\gamma>0$, then there exists positive constant $C_2$, independent of $\varepsilon$, and the solution of~\eqref{eq:nowerowneq} satisfies:}
	\begin{align}
		\left\|\xi(\varepsilon,t)\right\|&\leq \varepsilon C_1 \exp\left[-\left(\frac{1}{2}-\varepsilon C_1 C_2\right)\frac{t}{\varepsilon}\right]\nonumber\\
		\left\|\eta\right\|\cdot\left\|\xi(\varepsilon,t)\right\|^{-1}&\leq C_2,
	\end{align}
\textit{where $C_1=\sqrt{n+3}+\frac{1}{\omega_c}\sum_{j=1}^{n+2}\frac{L^j}{j!}\left\|\left(\bm{H}_q+\bm{I}_{n+3}\right)^{j^2}\right\|$ with $\bm{I}_{n+3}$ is the $n+3$ order identity matrix and:}
\begin{equation}
	\bm{H}_0 = \begin{bmatrix}
    7\omega_c^{6} & 0 & 0 & 0 & 0 & 0 & 0\\
		0 & 0 & 0 & 0 & 0 & 0 & 0\\
		0 & 0 & 0 & 0 & 0 & 0 & 0\\
		0 & 0 & 0 & -7\omega_c & 0 & 0 & 0\\
		0 & 0 & 0 & 0 & 0 & 0 & 0\\
		0 & 0 & 0 & 0 & 0 & 0 & 0\\
		0 & 0 & 0 & 0 & 0 & 0 & 0
	\end{bmatrix}.\nonumber
\end{equation}
\end{itemize}

Based on the standard result of the singular perturbation theory~\cite{Singular} as well as (i) of Theorem~1, one can conclude that (ii) holds if (i) of Theorem~1 holds. Also, it is stated in the above theorem that~\eqref{eq:nowerowneq} is exponentially stable and uniformly asymptotically stable if $\left\|\xi_0\right\|=O(\varepsilon)=O\left(\frac{1}{\omega_o}\right)$ and $\eta=\dot{F}$ exists.
\color{black}That directly implies that results of conventional ADRC stability analysis in~\cite{Gao-stabilitySingPert} can be straightforwardly extended to the proposed ADRC with RESO with the difference of having the error dynamic system~\eqref{eq:nowerowneq} with a more general coefficient matrix. To be specific, the coefficient matrix $\bar{\bm{A}}$ from~\eqref{eq:nowyyysystSS} can be replaced by a more general matrix, say $\tilde{\bm{A}}$, as long as $\tilde{\bm{A}}$ can lead to the matrix $\bm{H}_q$ in~\eqref{eq:omegahh} being decomposable into a form $\bm{H}_q=\bm{A}_q+\varepsilon^m\bm{\tilde{H}}_\varepsilon$, with $1\leq m<\infty$ and $\bm{\tilde{H}}$ being a square matrix of the same size as $\bm{H}_q$. For that reason, a detailed proof is omitted here in order to avoid redundancy of results.\color{black}


\color{black}

\section{Hardware validation}

\begin{figure}[t]
			\centering
				\includegraphics[width=0.75\textwidth]{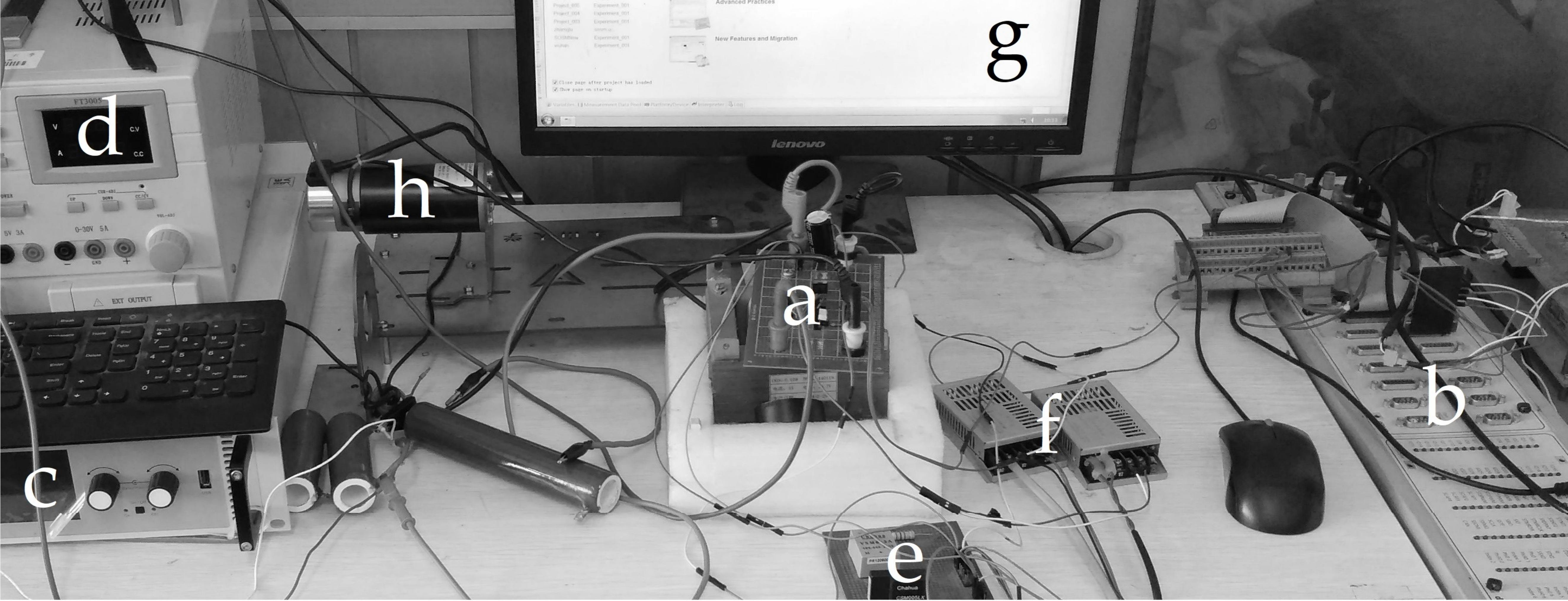}\vspace{0.5em}\\
				\includegraphics[width=0.75\textwidth]{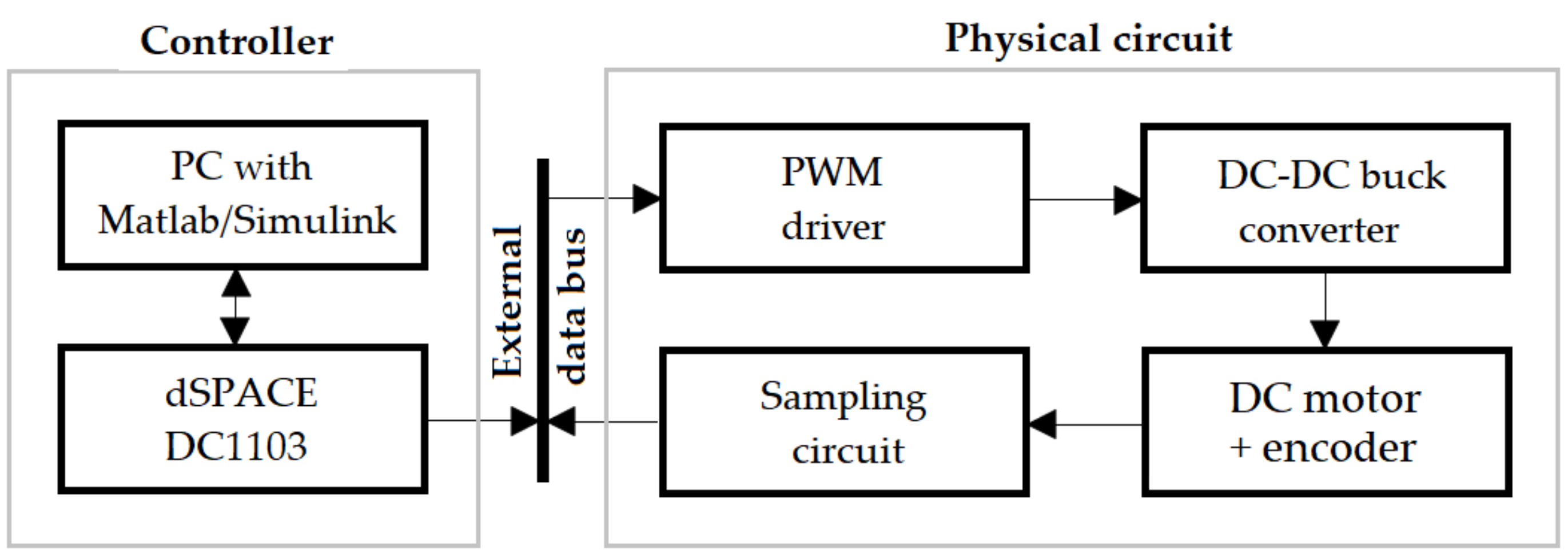}
		\caption{The DC-DC buck power converter-DC motor laboratory platform: with converter (a), real-time dSpace controller (b), external input voltage (c), digitial oscilloscope (d), voltage and current sensors (e), A/D converters (f), desktop computer with Matlab/Simulink software (g), and a DC motor with shaft-mounted incremental encoder with a tension controller for user-defined external disturbance generation (h).}
	\label{fig:platformConf}
\end{figure}

Three hardware experiments (E1-E3) were conducted on a real converter-fed motor laboratory platform (Fig.~\ref{fig:platformConf}) to validate the efficacy of the proposed control solution. The results of the proposed ADRC with RESO were quantitatively compared with the results obtained with ADRC with generalized proportional-integral observer (GPIO) and a standard PI controller. The ADRC with GPIO is a popular advanced motion controller that uses a higher-order polynomial representation of disturbances in case of systems subject to complex disturbances~\cite{SiraDCDCgpi}. The PI controller was selected since it represent the standard industrial control solution.

In order to allow a fair comparison between the two tested ADRC designs, the same tuning methodology from~\cite{Gao-pole-placement} (mentioned in Sect.~\ref{sec:Stabanaalal}) has been utilized for both of them. The gains were thus calculated upon~\eqref{eq:wspolpoly} and gathered in Table~\ref{tab:TunParam}. Choosing specific values of $\omega_c$, $\omega_o$ came from a compromise between control/estimation precision and amplification of sensor noise.

\begin{table}
\centering
\caption{Bandwidth-parametrization of observer gains $l_{1-7}$ and controller gains $k_{0-3}$ (same for RESO and GPIO, but with GPIO always having $\hat{\omega}_r=0$).}
\label{tab:TunParam}
 \begin{tabular}{| c | c |}
 \hline
 Parameter & Value \\ [0.5ex]
 \hline\hline
$l_1$ & $7\omega_o-k_3$  \\
 \hline
 $l_2$ & $21\omega_o^2-k_2-l_1k_3-\hat{\omega}_r^2$  \\
 \hline
 $l_3$ & $35\omega_o^3-k_1-l_1k_2-l_2k_3-\hat{\omega}_r^2(l_1+k_3)$  \\
 \hline
$l_4$ & $35\omega_o^4-l_1k_1-l_2k_2-l_3k_3-\hat{\omega}_r^2(l_1k_3+l_2+k_2)$  \\
 \hline
$l_5$ & $21\omega_o^5-\hat{\omega}_r^2(l_3+k_1+l_1k_2+l_2k_3)$  \\
 \hline
$l_6$ & $7\omega_o^6-\hat{\omega}_r^2(l_4+l_1k_1+l_2k_2+l_3k_3)$ \\
 \hline
$l_7$ & $\omega_o^7-l_5\hat{\omega}_r^2$ \\
 \hline
 $k_0$ & $\omega_c^4$ \\
 \hline
 $k_1$ & $4\omega_c^3$ \\
 \hline
$k_2$ & $6\omega_c^2$ \\
 \hline
 $k_3$ & $4\omega_c$ \\
 \hline
  \end{tabular}
\end{table}





\subsection{Establishing nominal performance (E1)}

Test (E1) was about establishing a base for fair tests later as well as showing nominal performance of each tested control algorithms. The reference signal was set as:
\begin{equation}
	\omega_d\text{[rad/s]} =
		\begin{cases}
			100, t\in[0,1), \\
			200, t\in[1,2), \\
			300, t\in[2,3), \\
			400, t\in[3,4], \\
		\end{cases}
\end{equation}
and additionally filtered with a stable dynamics $H(s)=1/(0.025s^2+0.6s+1)$ to satisfy assumption A3. The three control structures were tuned to give comparable results of quality of tracking, signal profiles, and level of sensor noise. The order of GPIO was chosen to match with the order of RESO (7th order) thus making the observer gains of both observers with similar magnitudes. Both ADRC structures were tuned with: $\omega_c=0.35$, $\omega_o=140$, and the system gain in both cases was calculated upon the available plant model information as $\hat{b}_0=4.3015\times 10^{12}$. Both GPIO and RESO were designed as in~\eqref{eq:RESOoo} but with GPIO lacking the internal harmonic model (i.e. $\hat{\omega}_r=0$). Furthermore, the same form of control action~\eqref{eq:MMu} has been used in both ADRC cases however with different observers working to reconstruct total disturbance~$\hat{F}$.

For the PI controller design ($u(t)\triangleq k_pe(t)+k_d\int e(t)dt$), the gains were selected based on trial and error approach as $k_p=0.01$ and $k_i=0.25$, respectively. The derivative part had to be excluded from the design due to assumptions A2 and~A4.

\begin{figure}[p]
    \centering
    \begin{subfigure}[b]{0.49\textwidth}
        \includegraphics[width=\textwidth,page=1]{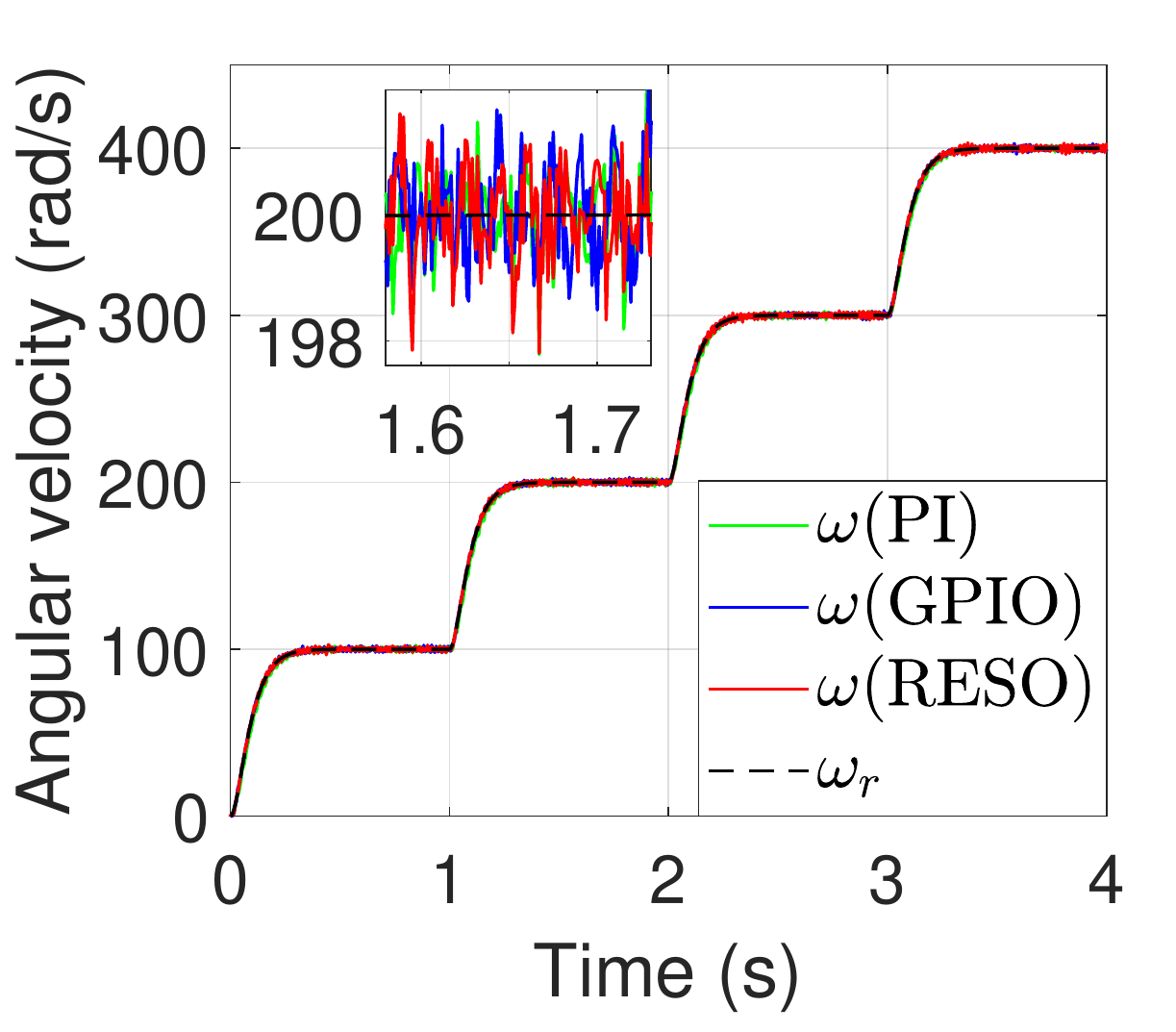}
    \end{subfigure}
    \begin{subfigure}[b]{0.49\textwidth}
        \includegraphics[width=\textwidth,page=2]{exp/exp1ALLMOD}
    \end{subfigure}
    \\
            \begin{subfigure}[b]{0.49\textwidth}
        \includegraphics[width=\textwidth]{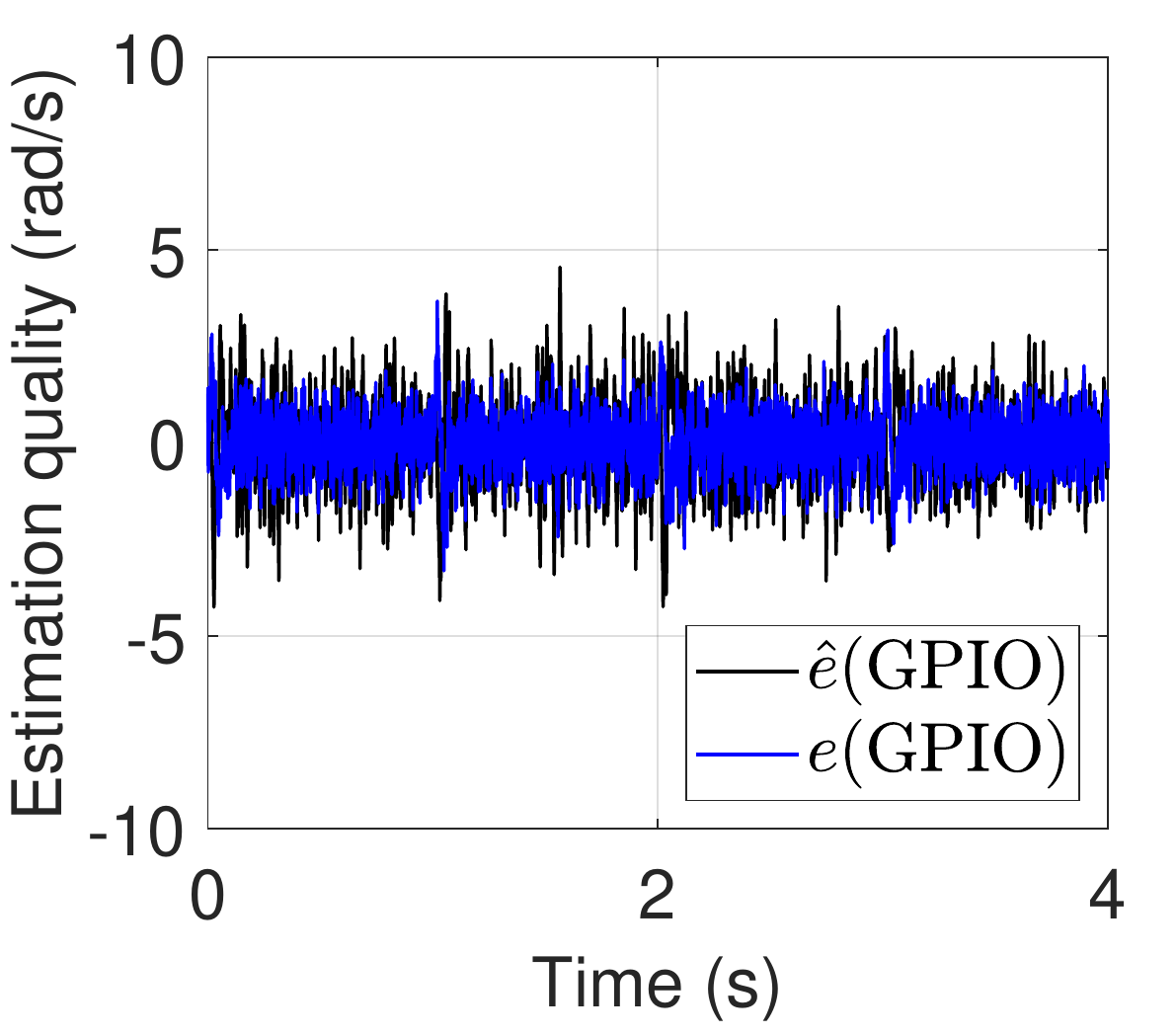}
				\end{subfigure}
    \begin{subfigure}[b]{0.49\textwidth}
        \includegraphics[width=\textwidth]{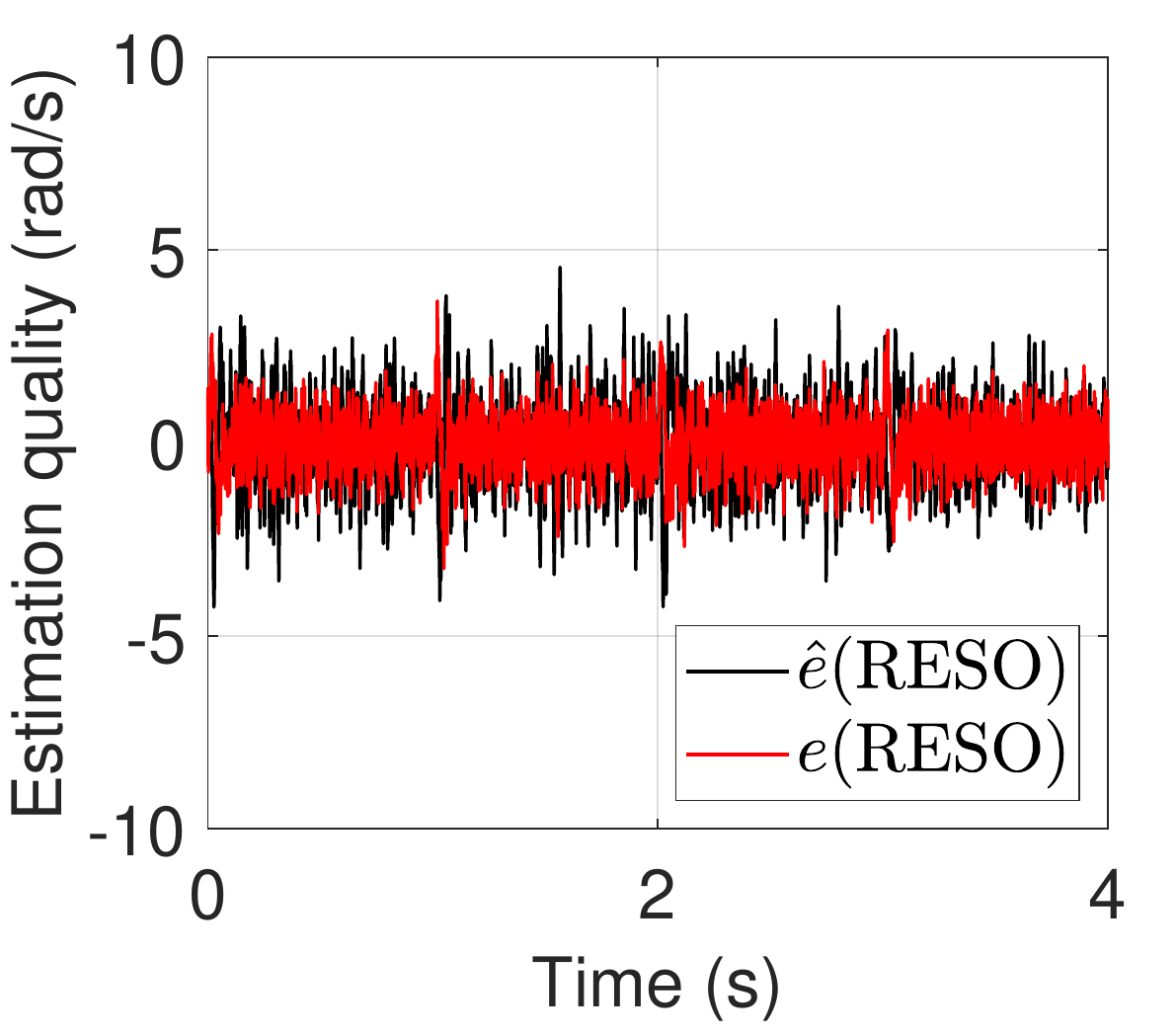}
    \end{subfigure}\\
    \includegraphics[width=0.49\textwidth]{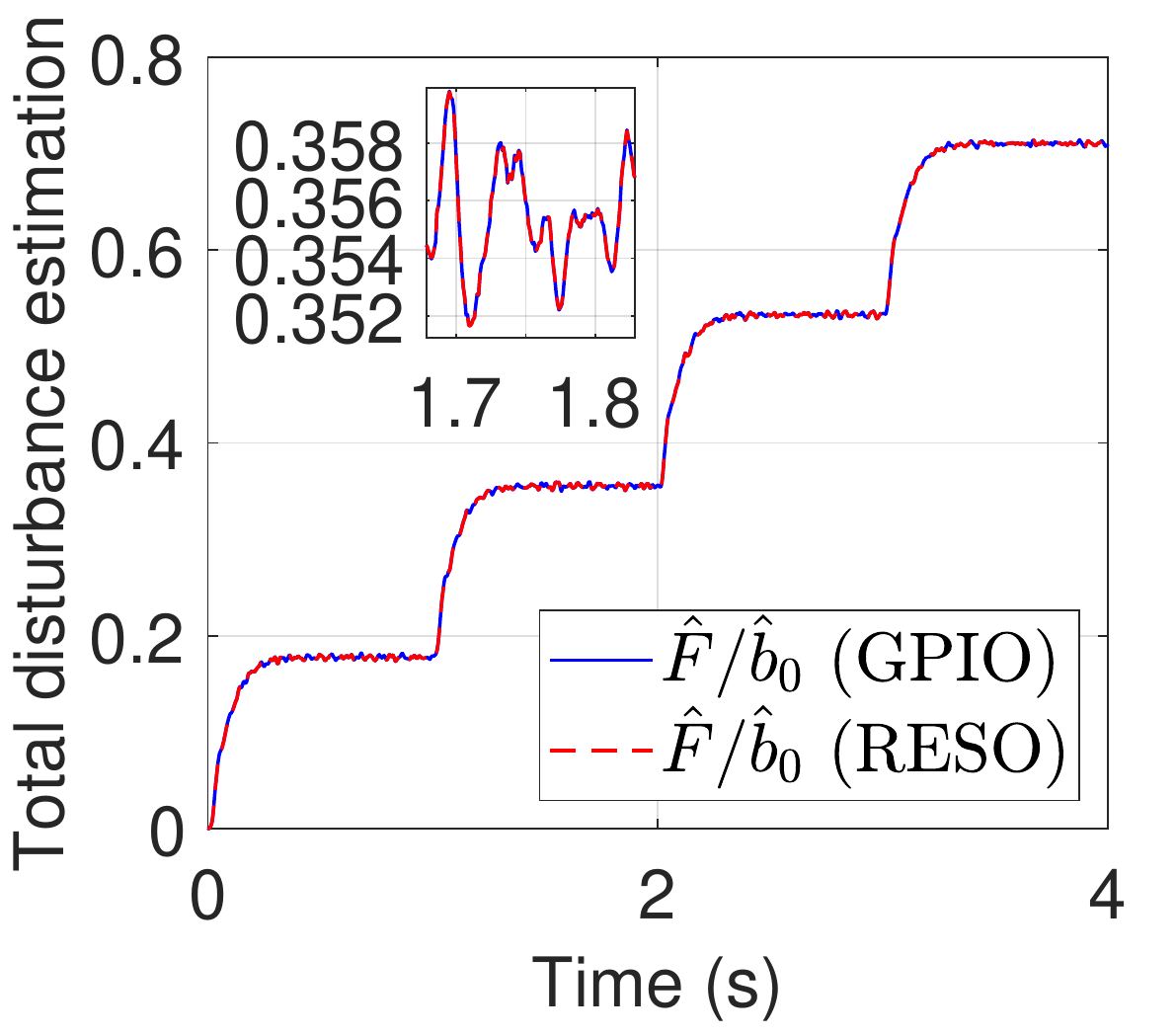}
    \caption{[E1] Performance of  PI control (\textcolor[rgb]{0,0.58,0}{green}), ADRC with GPIO (\textcolor[rgb]{0,0,1}{blue}), and proposed ADRC with RESO (\textcolor[rgb]{1,0,0}{red}).}
		\label{Exp1}
\end{figure}

The outcomes of E1 are shown in Fig.~\ref{Exp1}. The three considered control solutions give comparable quality of tracking ($\omega\rightarrow\omega_d$), level of measurement noise, and energy usage ($u$). The achieved behavior of each control algorithm is treated as a baseline for next tests.

\subsection{Robustness against load torque (E2)}

The robustness of the considered control methods was tested against a step-like (E2a) and a sinusoidal-like load torque disturbance (E2b), both generated using a tension controller (Fig.~\ref{fig:platformConf}) and applied to the system at $t=1$s. The step signal (E2a) had amplitude $c_0=1$Nm and the sinusoidal signal (E2b) had amplitude $a_1=1.35$Nm, $a_2=0$Nm, and frequency $\omega_r=6\pi$. In case E2b, $\hat{\omega}_r=\omega_r$ was set in case of RESO and $\hat{\omega}_r=0$ in case of GPIO.

\begin{figure}[t]
\centering
    \begin{subfigure}[b]{0.49\textwidth}
        \includegraphics[width=\textwidth]{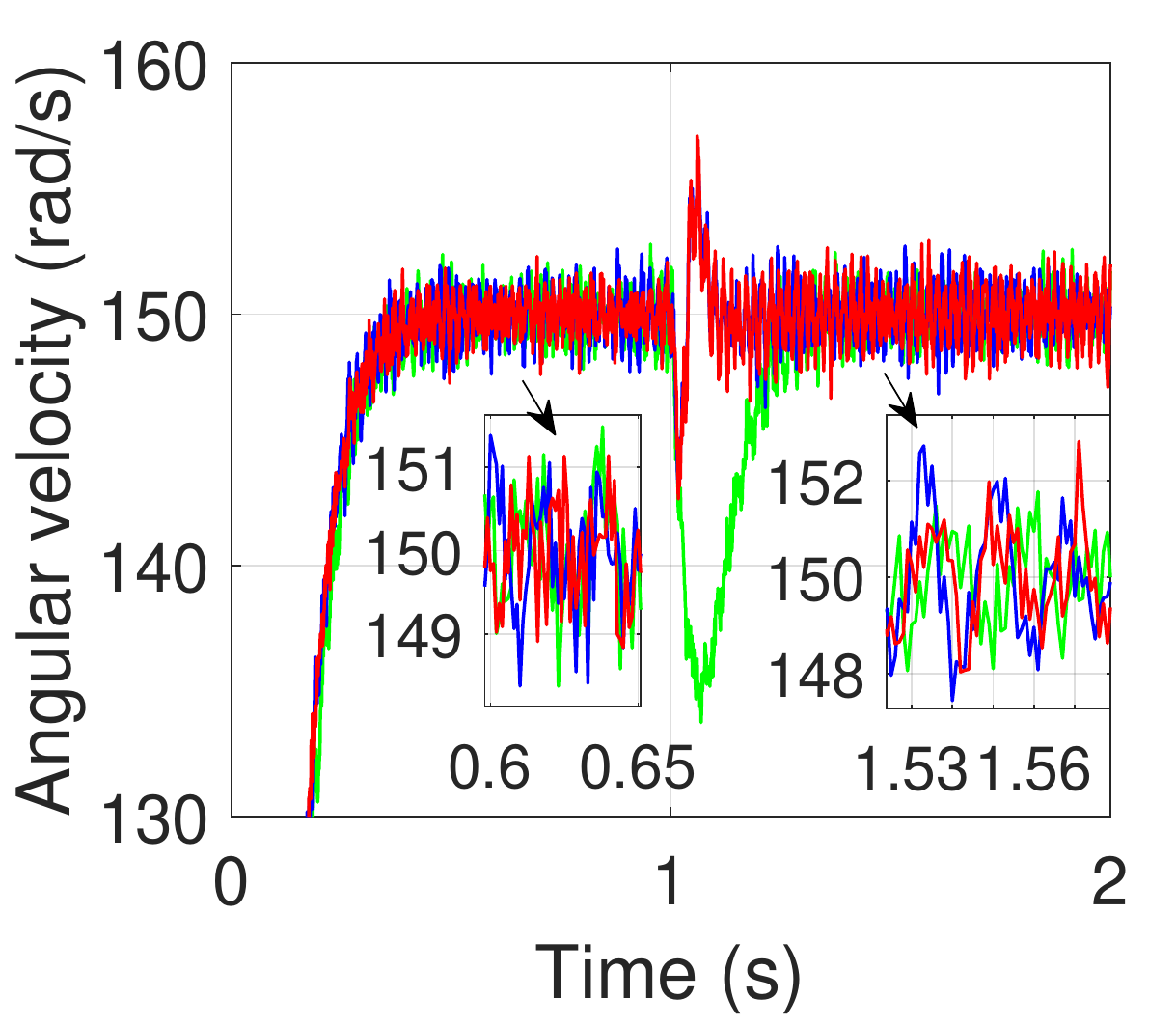}
				\end{subfigure}
    \begin{subfigure}[b]{0.49\textwidth}
        \includegraphics[width=\textwidth]{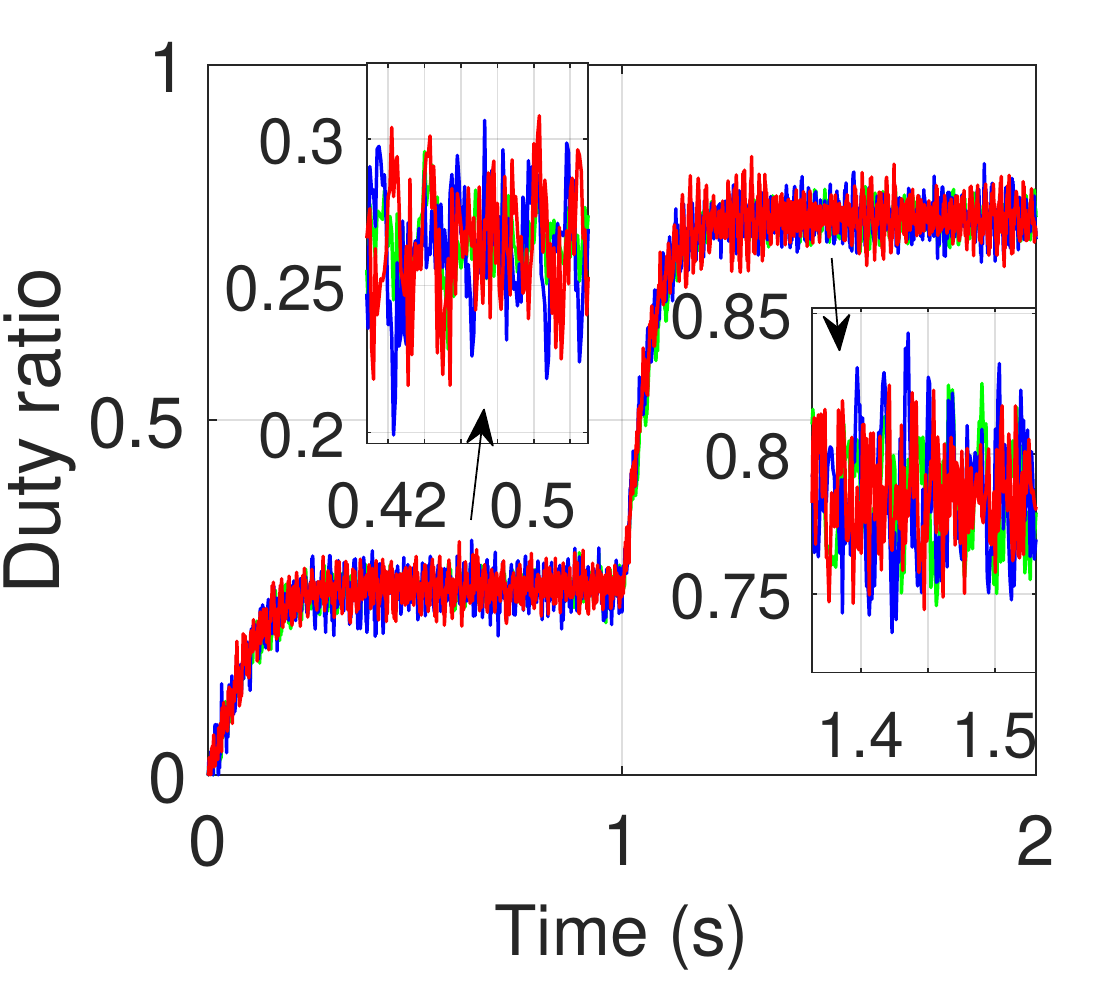}
    \end{subfigure}\\
    \includegraphics[width=0.49\textwidth]{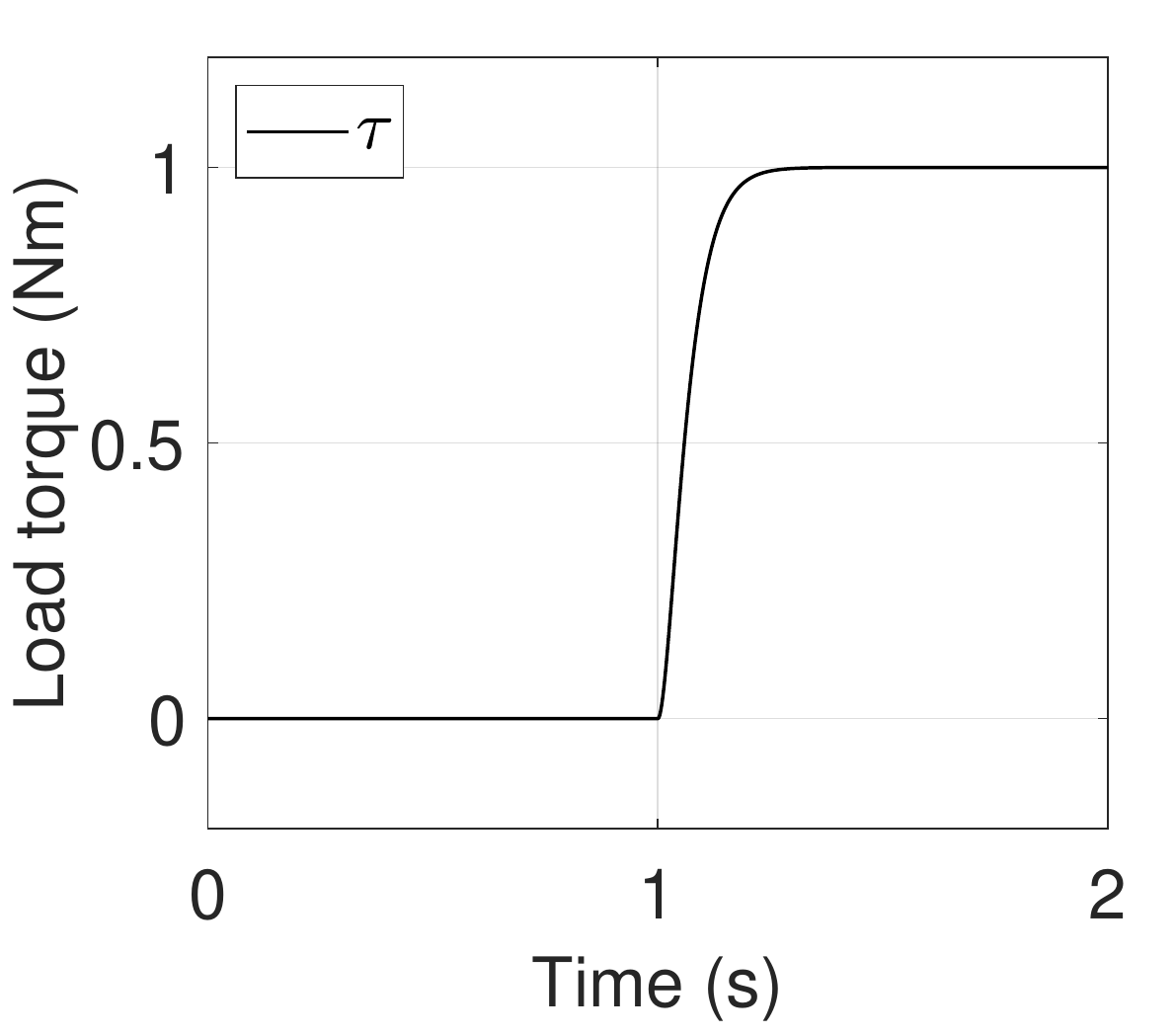}
    \caption{[E2a] Performance of PI control (\textcolor[rgb]{0,0.58,0}{green}), ADRC with GPIO (\textcolor[rgb]{0,0,1}{blue}), and proposed ADRC with RESO (\textcolor[rgb]{1,0,0}{red}) under the effect of load torque (black).}\label{Exp2a}
\end{figure}

The outcomes E2a are seen in Fig.~\ref{Exp2a}. The unmodeled step disturbance resulted in a significant $15$rad/s velocity drop in case of PI, which took it 0.38s to recover to nominal operation. The added disturbance also affected both ADRCs but thanks to the constant disturbance model ($F_p$) in both GPIO and RESO designs, the change was not that significant. In both cases, the step disturbance was timely estimated and mitigated, resulting in around $0.14$s recovery time.

\begin{figure}[t]
\centering
    \begin{subfigure}[b]{0.49\textwidth}
        \includegraphics[width=\textwidth]{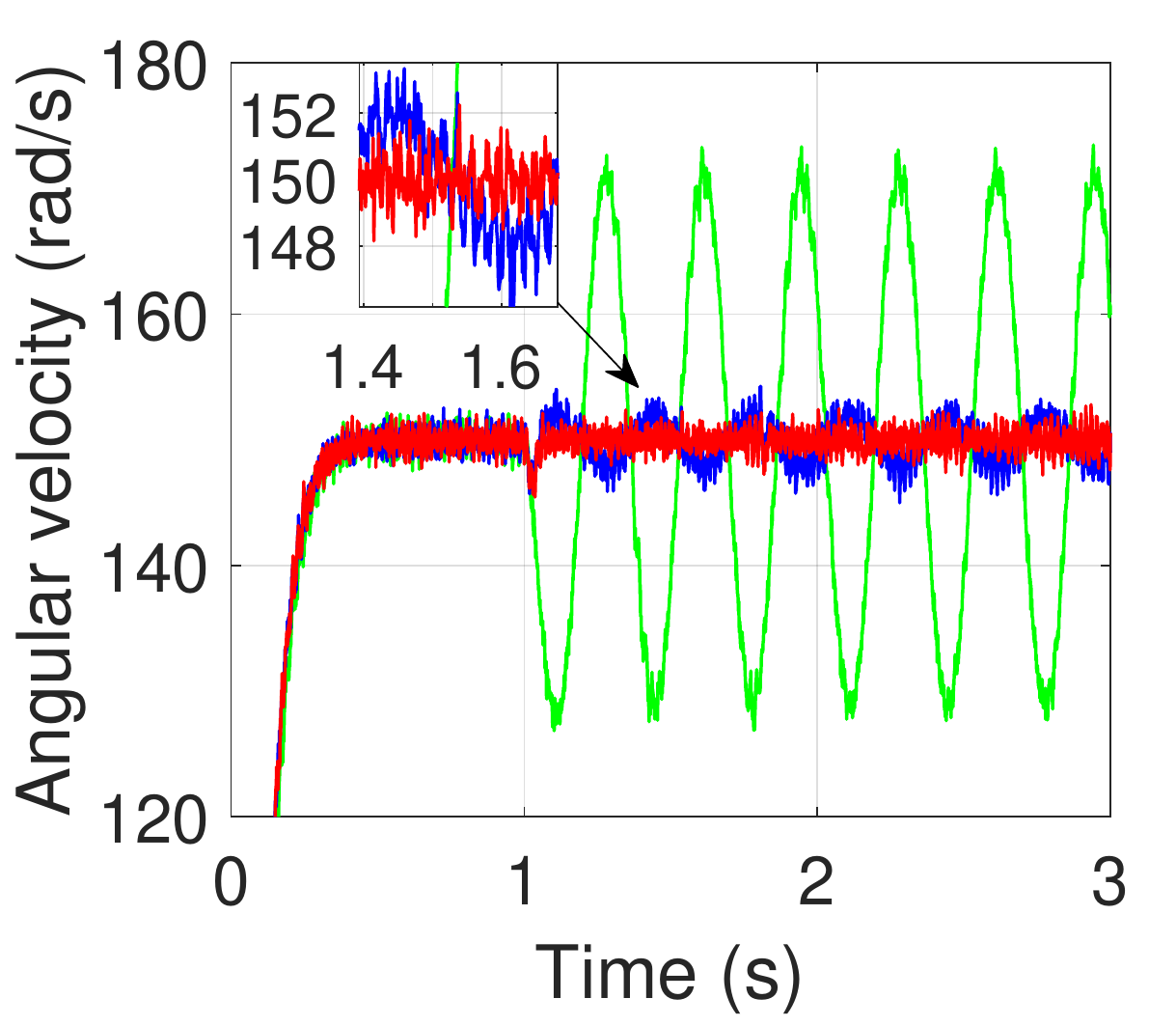}
    \end{subfigure}
    \begin{subfigure}[b]{0.49\textwidth}
        \includegraphics[width=\textwidth]{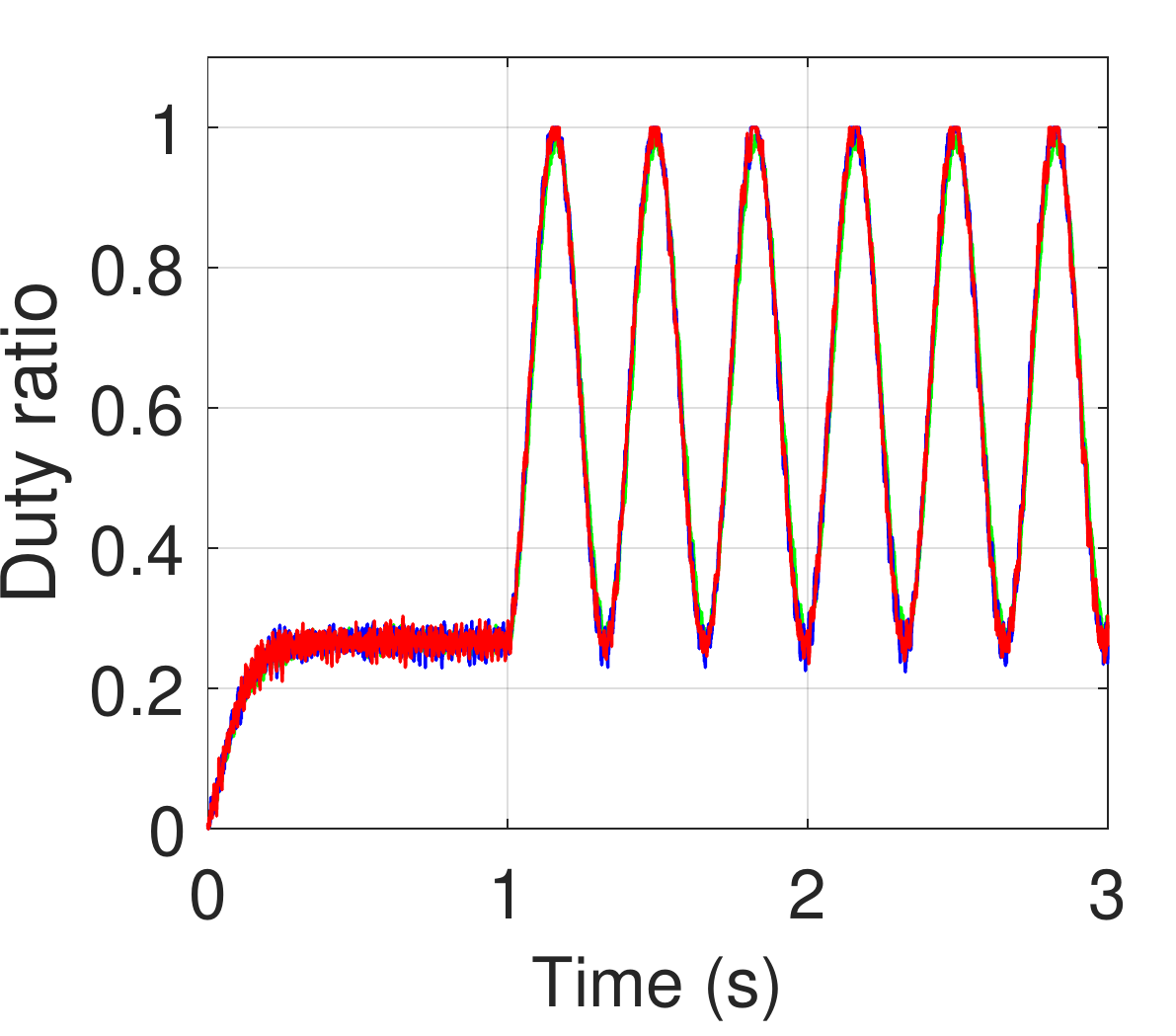}
    \end{subfigure}
    \begin{subfigure}[b]{0.49\textwidth}
        \includegraphics[width=\textwidth]{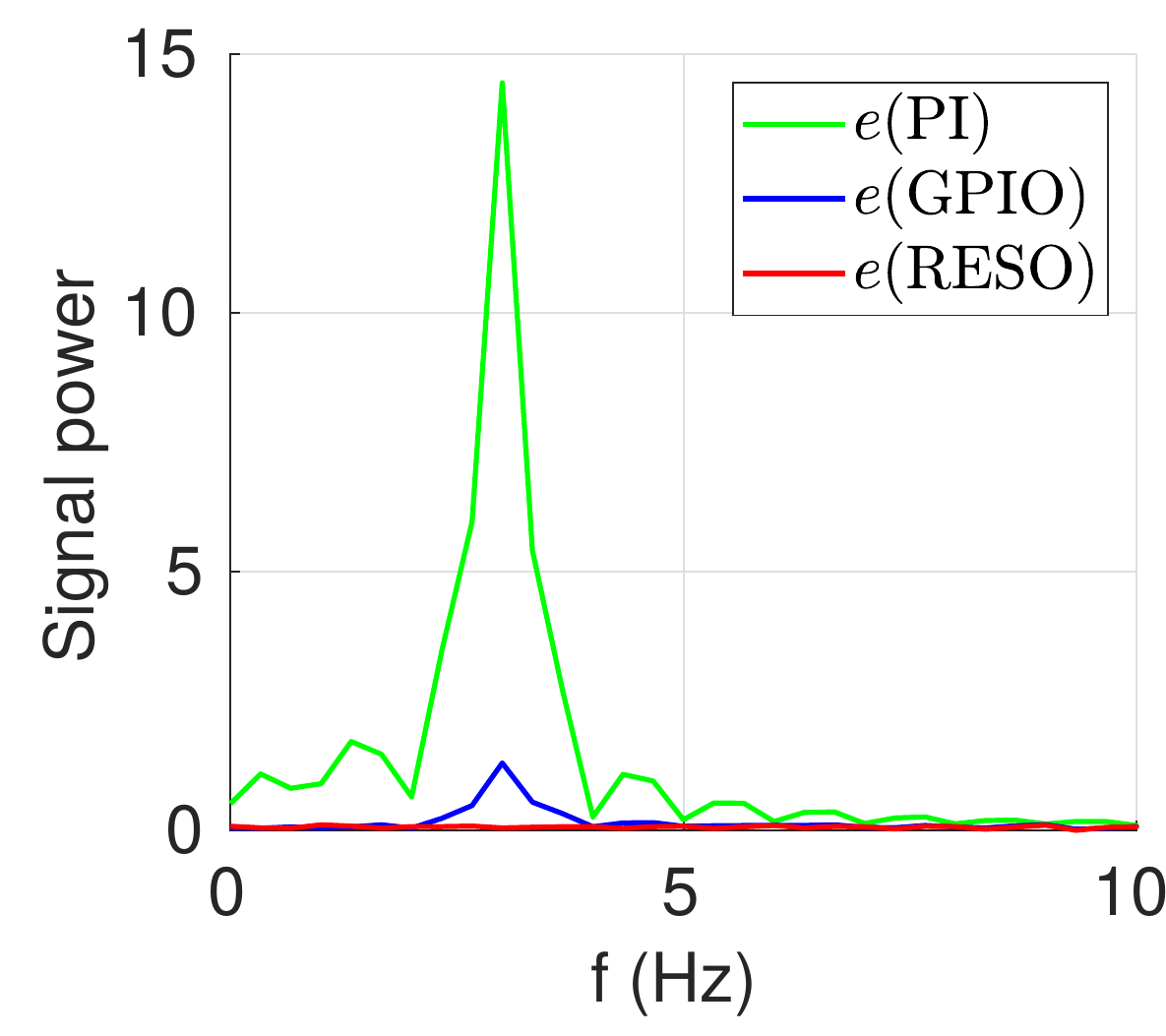}
    \end{subfigure}
    \begin{subfigure}[b]{0.49\textwidth}
        \includegraphics[width=\textwidth]{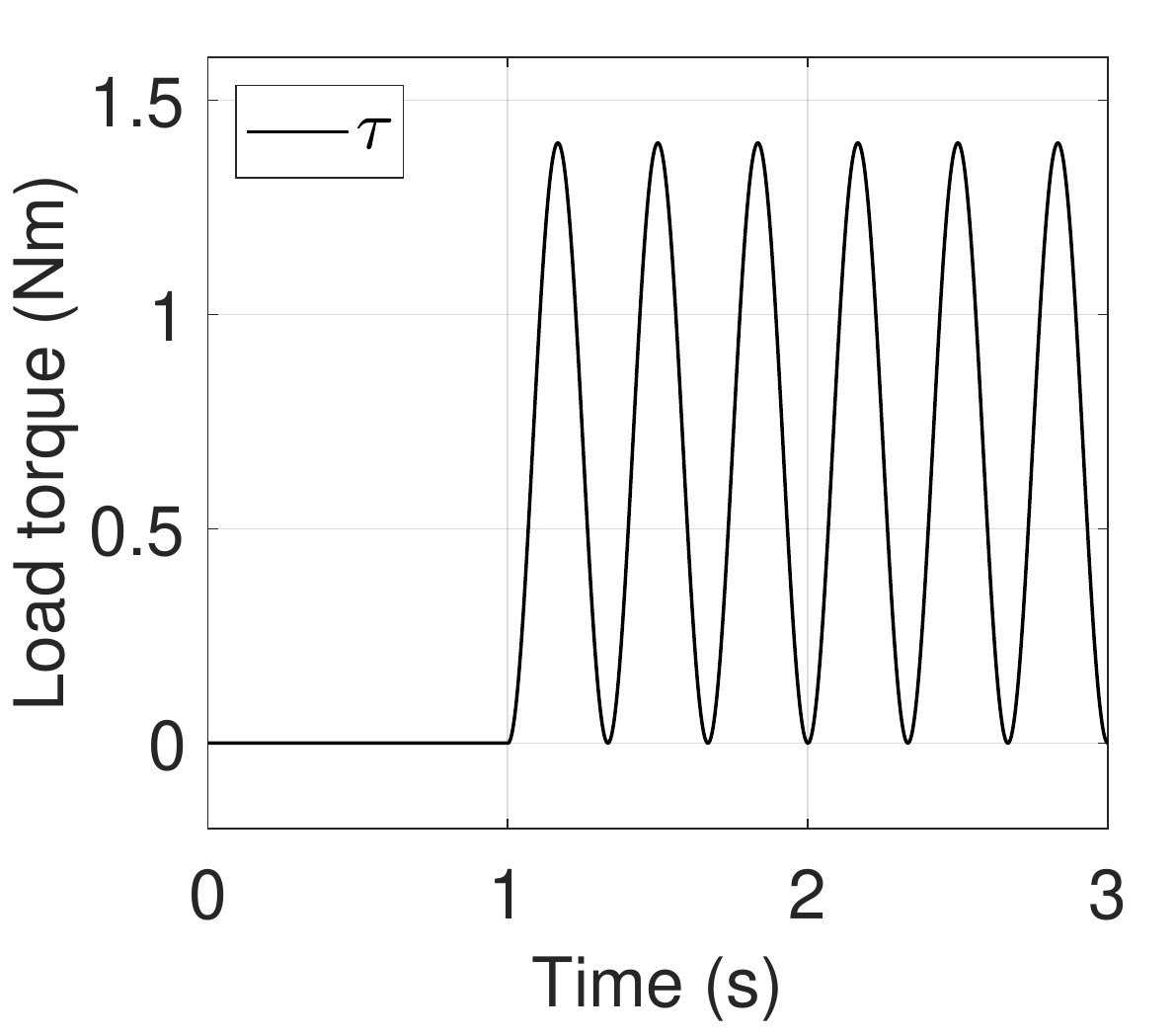}
    \end{subfigure}
    \caption{[E2b] Performance of PI control (\textcolor[rgb]{0,0.58,0}{green}), ADRC with GPIO (\textcolor[rgb]{0,0,1}{blue}), and proposed ADRC with RESO (\textcolor[rgb]{1,0,0}{red}) under the effect of load torque (black).}
		\label{Exp2b}
\end{figure}

The outcomes E2b are seen in Fig.~\ref{Exp2b}. The unmodeled harmonic load torque disturbance significantly influenced the standard PI controller causing a $\pm22.4$rad/s output oscillations. The test has shed a light on the structural limitation of the GPIO-based ADRC. The exclusive use of constant disturbance model ($F_p$) in GPIO allows only partial reconstruction of fast-varying sinusoidal signal. Improving quality of disturbance estimation would require higher-order GPIO and/or higher gains - both posing certain practical problems, like sensor noise over-amplification. The proposed ADRC with RESO managed to attenuate the sinusoidal disturbance and almost entirely keep its nominal performance thanks to the built-in model of harmonic disturbance.


\subsection{Robustness against parametric uncertainty (E3)}

The experiment E3 was about verifying what is the level of modeling discrepancy in selecting the estimated resonant frequency ($\hat{\omega}_r$) that the proposed ADRC with RESO can tolerate. This test was inspired by control practice in which often times $\hat{\omega}_r\neq\omega_r$. For this test, the frequency of the harmonic disturbance was set to $\omega_r=6\pi$.

\begin{figure}[p]
\centering
    \begin{subfigure}[b]{0.49\textwidth}
        \includegraphics[width=\textwidth,page=1]{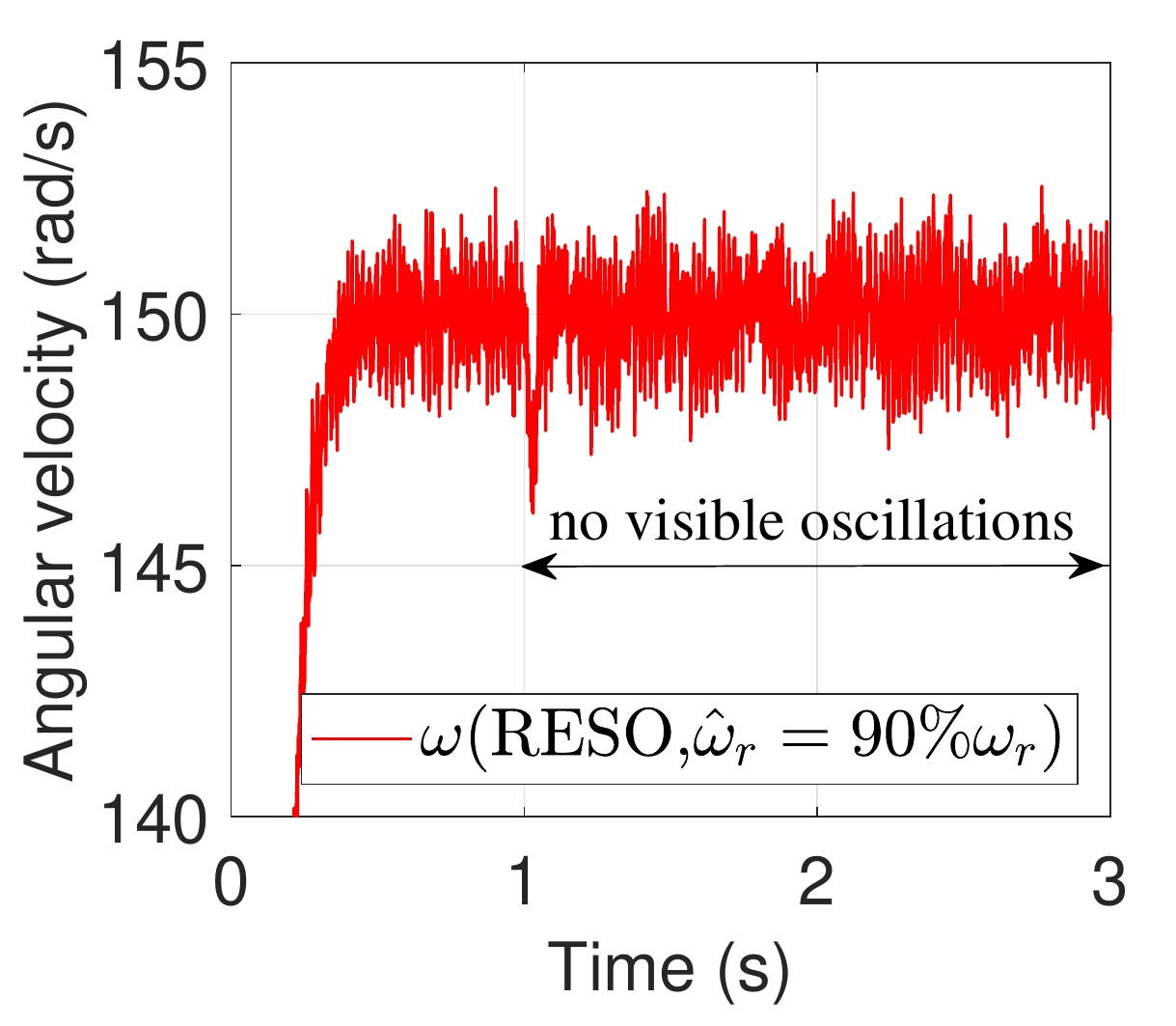}
    \end{subfigure}
		\begin{subfigure}[b]{0.49\textwidth}
        \includegraphics[width=\textwidth,page=4]{exp/exp4ALLMOD}
    \end{subfigure}
		\\
    \begin{subfigure}[b]{0.49\textwidth}
        \includegraphics[width=\textwidth,page=6]{exp/exp4ALLMOD}
    \end{subfigure}
    \begin{subfigure}[b]{0.49\textwidth}
       \includegraphics[width=\textwidth,page=7]{exp/exp4ALLMOD}
    \end{subfigure}
    \begin{subfigure}[b]{0.49\textwidth}
        \includegraphics[width=\textwidth]{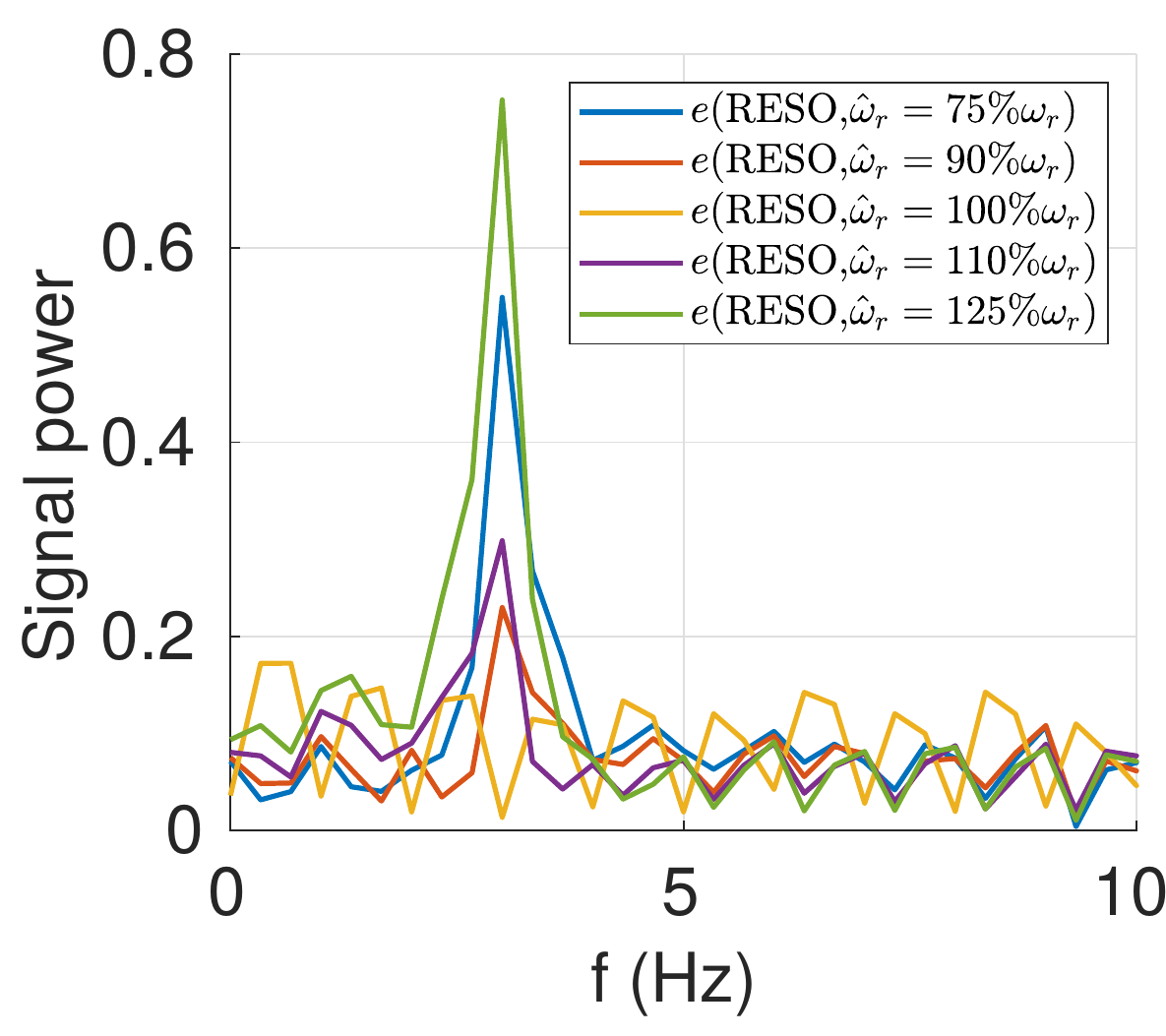}
    \end{subfigure}
    \begin{subfigure}[b]{0.49\textwidth}
        \includegraphics[width=\textwidth]{exp/Exp3extDist}
    \end{subfigure}
		 \caption{[E3] Parametric robustness of ADRC with RESO under the effect of load torque.}
		\label{Exp3}
\end{figure}

The outcomes of E3 are seen in Fig.~\ref{Exp3}. It can be seen that, due to the disturbance-centered design, the proposed ADRC with RESO has certain robustness against inaccuracy of $\hat{\omega}_r$ (top row). As expected, and confirmed with middle row plots, this robustness is limited. When the modeling discrepancy reaches up to $\pm25\%$ of the actual value, the ADRC with RESO began to have visible oscillations, just like GPIO in test E2b. This implies that the ADRC with RESO is suitable to control scenarios which have known, or "roughly" known, frequency of the harmonic disturbance. A systematic, analytical method of establishing robustness bounds is difficult in this case and is yet to be developed.

\section{Conclusions}

An advanced motion control solution for power converter-fed DC motors has been proposed. The utilization of a resonant extended state observer, working under the framework of active disturbance rejection control, allowed to enhance precision of angular velocity tracking and its robustness against even complex, harmonic disturbances. At the same time, the proposed control algorithm was shown to be straightforward to implement in practice and to have similar level of energy consumption with some standard methods. The claims have been supported with experimental results and a rigorous stability analysis.


In the future work, the practical part can be realized using a series excitation motor because of the limitation of permanent magnet DC motor to non-industrial applications. Also, a cascade control structure can be introduced with inner, current feedback loop in order to increase circuit safety. Additionally, a frequency estimator method can be combined with RESO to provide on-line information about the harmonic disturbance frequency.

\section*{Acknowledgment}

The Authors would like to thank students Han Wu and Zhang Lu for their help with the experimental part of the paper.

\section*{Conflict of interest statement}

Conflict of interest - none declared


\bibliography{mybibfileRM}

\end{document}